\documentstyle[12pt]{article}
\newcommand{\newsection}{    
\setcounter{equation}{0}
\section}
\def\appendix#1{
\addtocounter{section}{1}
\setcounter{equation}{0}
\renewcommand{\thesection}{\Alph{section}}
\section*{Appendix \thesection\protect\indent #1}
\addcontentsline{toc}{section}{Appendix \thesection\ \ \ #1}
}
\newcommand{\rf}[1]{(\ref{#1})}
\newcommand{\eq}[1]{Eq.~(\ref{#1})}

\def\be{\begin{equation}}
\def\ee{\end{equation}}
\newcommand{\beq}{\begin{equation}}
\newcommand{\eeq}{\end{equation}}
\newcommand{\bea}{\begin{eqnarray}}
\newcommand{\eea}{\end{eqnarray}}
\newcommand{\g}{\gamma}
\renewcommand{\l}{\lambda}
\renewcommand{\b}{\beta}
\renewcommand{\a}{\alpha}
\newcommand{\baA}{\xi}

\newcommand{\om}{\omega}

\newcommand{\oh}{\frac{1}{2}}

\newcommand{\dg}{\dagger}

\newcommand{\tr}{{\,\rm Tr}\;}
\newcommand{\str}{{\,\rm tr}\;}
\newcommand{\cM}{{\cal M}}
\newcommand{\cL}{{\cal L}}

\newcommand{\cim}{\oint_{C}\frac{d\omega}{2\pi i}}
\newcommand{\cimc}{\oint_{P}\frac{d\omega}{4\pi i}}
\newcommand{\ci}{\oint_{C}\frac{d\omega}{2\pi i}\;\frac{
V'(\omega)}{p-\omega}}
\newcommand{\cic}{\oint_{P}\frac{d\omega}{4\pi i}\;\frac{
V_C'(\omega)\,\om}{p^2-\omega^2}}
\newcommand{\lL}{\lambda}
\newcommand{\LL}{\Lambda}
\newcommand{\h}{\eta}

\newcommand{\eps}{\varepsilon}

\newcommand{\LA}{\left\langle}
\newcommand{\RA}{\right\rangle}
\newcommand{\la}{\left\langle}
\def\e{{\,\rm e}\,}
\newcommand{\ie}{{\it i.e.}\ }

\newcommand{\half}{{\textstyle{1\over 2}}}
\renewcommand{\d}{{{\partial}}}

\newcommand{\ra}{\rightarrow}
\def\hm{the hermitian matrix model}
\def\dsl{the double scaling limit}

\begin{document}
\topmargin 0pt
\oddsidemargin 5mm
\headheight 0pt
\headsep 0pt
\topskip 9mm

\hfill    NBI-HE-92-89

\hfill PAR--LPTHE--92-49

\hfill FTUAM 92-44


\hfill November 1992

\addtolength{\baselineskip}{0.20\baselineskip}

\begin{center}

\vspace{24pt}
{\large \bf
Matrix model calculations beyond the spherical limit
}

\vspace{24pt}

{\sl J. Ambj\o rn }

\vspace{6pt}

 The Niels Bohr Institute\\
Blegdamsvej 17, DK-2100 Copenhagen \O , Denmark\\

\vspace{12pt}

{\sl L.\ Chekhov}\footnote{
Permanent address: Steklov Mathematical Institute,
Vavilov st.42, GSP-1 117966 Moscow, Russia}

\vspace{6pt}

L.P.T.H.E., Universit\`e Pierre et Marie Curie \\
4, pl. Jussieu, 75252, Paris Cedex 05, France\\

\vspace{12pt}

{\sl C.\ F.\ Kristjansen}

\vspace{6pt}

 The Niels Bohr Institute\\
Blegdamsvej 17, DK-2100 Copenhagen \O , Denmark\\

\vspace{12pt}

{\sl Yu.\ Makeenko}

\vspace{6pt}

 Institute of Theoretical and Experimental Physics\\
 B.Cheremushkinskaya 25, 117259 Moscow, Russia\\

\end{center}

\vspace{18pt}

\begin{center}
{\bf Abstract}
\end{center}
We propose an improved iterative scheme for calculating higher genus
contributions to the multi-loop (or multi-point) correlators and the
partition function of the hermitian one matrix model.
We present explicit results up to genus two. We develop
a version which gives directly the result in the double scaling limit and
present explicit results up to genus four.
Using the latter version we prove that the hermitian and the
complex matrix model are equivalent in the double scaling limit and
that in this limit they are both equivalent to the
Kontsevich model.
We discuss how our results away from the double scaling limit are
related to the structure of moduli space.

\newpage

\newsection{Introduction}
The hermitian one matrix model is interesting for many reasons.
The matrices involved being $N\times N$, the model can be solved
in the limit when $N\ra\infty$, as
 was first demonstrated in Ref.~\cite{BIPZ78}. Furthermore the model
admits a $1/N^2$-expansion.
This expansion is the simplest example of the \mbox{'t Hooft} topological
(or genus) expansion \cite{Hoo74} --- in the given case of a $D=0$
 dimensional quantum field theory. If the leading order (planar or genus zero
approximation) result for some quantity, say the vacuum energy, is normalized
to be of order $1$, the genus $g$ contribution is of order
${\cal O}(N^{-2g})$.
Even though the hermitian matrix model is often considered a toy model,
its perturbative expansion
reveals the combinatorial problems one will encounter in the
multi-dimensional case: The sum of graphs contributing to a given order of
the genus expansion is convergent, while the attempt of summing over all
genera leads to a divergence.
This was one of the original motivations for studying the genus expansion
of \hm.

One more application of \hm\ is due to the fact that it
describes~\cite{DyTriang} discretized 2-dimensional random surfaces
and hence 2D quantum gravity.
In this language  a diagram of genus $g$ is equivalent to
a (piecewise linear) surface of genus $g$.
The hermitian one matrix model was extensively studied from this point of
view after the observation~\cite{Kaz89} that it might not only describe
pure 2D gravity but also some minimal conformal field theories coupled
to 2D quantum gravity\footnote{Although the relation conjectured in
\cite{Kaz89} turned out to be wrong, the paper was nevertheless
instrumental for the later developments.}. The relation given in
\cite{Kaz89} for genus zero
was extended to any order of the genus
expansion in the so-called double scaling limit,
where the coupling constants
approach some critical values in a $N$-dependent way~\cite{dsl}.

In the double scaling limit many interesting approaches have been developed
in order to deal with higher genus contributions.
The technique of exactly integrable systems (KdV and KP hierarchies)
was applied to the model in Ref.~\cite{GM90b}.
It provides an algorithm for genus by genus
calculations in \dsl. Another approach proposed by Witten~\cite{W1} is based
on the interpretation of \dsl\ of the matrix model
as a topological theory. The problem is then reduced
to calculating intersection indices on  moduli space. In this approach
it was possible to obtain explicit results in
\dsl\ up to genus four~\cite{Hor90}. A further
development of this approach was the representation~\cite{Kon91} of the
 generating function
for the intersection indices as a hermitian one matrix model in an external
field (the Kontsevich model).

For \hm\ away from \dsl, the original genus zero solution~\cite{BIPZ78} was
 obtained by solving the large-N integral saddle point equation for the
spectral
density. To extend the results to higher genera, the orthogonal polynomial
technique was applied in Ref.~\cite{Bes79}. While this method works
well for the partition function in the case of simple potentials
and lower genera, it is difficult to implement if one wants an
explicit calculation of multi-point correlators and if the potential
is ``generic''.

Another method of solving \hm\ is based on the loop equations~\cite{Mig83}
which were first written down for \hm\ in Ref.~\cite{Wad81}.
In this approach one solves the genus zero equation and uses
the solution in an iterative genus by genus procedure.
In the scheme originally
proposed in Ref.~\cite{Mig83} and elaborated in Ref.~\cite{Dav90},
one determines the higher genera contributions by solving algebraically
the higher loop equations and imposing on all correlators the same
analyticity structure as that of their genus zero versions.
Following this line of action
genus one correlators for  $\phi^4$ and $\phi^6$
potentials were calculated in ref.~\cite{jm}.
However, this method involves the entire chain of loop equations and is
in practise only applicable when one only wants to calculate a few terms
in the genus expansion for a potential with a small number of terms.

An alternative iterative procedure for calculating higher genus
contributions to the correlators of the
\hm\ has been proposed recently~\cite{ACM92}. It is based only on the
first in the chain of loop equations, is applicable for
an arbitrary potential and can be pursued order by order of the genus
expansion without any problems.
A key point in this scheme is a change of variables to the so called
{\it moments} of
the potential. The scheme is  close in spirit to the iterative
solution of the unitary and hermitian (with cubic potential) one matrix
models in an external field~\cite{GN91}%
\footnote{ For the Kontsevich model similar moments has been used in
Ref.~\cite{IZ}.}.
The genus one
partition function and correlators were calculated  for an even
potential in Ref.~\cite{cm} and for an arbitrary potential
 in Ref.~\cite{ACM92}, where it was also
 proven that genus $g$ contribution to the \mbox{$n$-loop} correlator depends
on at most  $2\times(3g-2+s)$ lower moments. An analogous algorithm for
the complex matrix model was developed in Ref.~\cite{complex} and the genus
$2$ and $3$ correlators were explicitly  calculated for an arbitrary potential.

In the present paper we  propose an improved algorithm for an iterative
calculation of higher genus contributions to the multi-loop correlators
and to the partition function of
the hermitian one matrix model.
We perform calculations
up to and including genus two.
Our algorithm differs from that of Ref.~\cite{ACM92} by
a redefinition of the moments,
which has the advantage that it allows us to
develop a version which gives directly the result in the double scaling limit
where we perform explicit calculations up to  and including genus four.
We compare the analysis to that of the complex matrix model and prove
in particular that the hermitian and
complex matrix models are equivalent in the double scaling limit.
We formulate a limiting procedure which allows us to
obtain the Kontsevich from the hermitian matrix model and use it to
prove in a very direct
manner the equivalence between the double scaling limit of
the hermitian matrix model and the Kontsevich model.

We discuss a possible interpretation of our results for \hm\
as representing a particular discretization of moduli space
which preserves basic geometrical properties
such as intersection indices, and which allows for a detailed study of
the {\it boundary} of moduli space \cite{Ch}.
We conjecture that the moment-technique developed in the present paper
might be an  efficient tool in the study of this ``fine structure''
of moduli space.

\vspace{12pt}

The paper is organized as follows:
In Section~2 we introduce the basic ingredients of the iterative
procedure, the
loop insertion
operator, the loop equation and the new variables --- the moments.
In Section~3 we describe the iterative procedure and present the results for
the average of loop operators and the partition function
for \hm\ up to genus two.
In Section~4 we demonstrate in detail how our iterative procedure works in
the double scaling limit and calculate the correlators and the partition
function
up to genus four.
In Section~5 we show that the Kontsevich model can be obtained as a
certain limit of the hermitian matrix model,
prove the equivalence of the
double scaling limit of the hermitian matrix model and the Kontsevich
model and discuss the possible connection between our results for \hm\
away from the double scaling limit
and the structure of the  discretized moduli space. Finally Section 6
contains a short discussion of future perspectives and
in the Appendix we develop  a double scaling version of the iterative
procedure for the complex matrix model proposed in \cite{complex},
and prove that the double scaling  limits of the complex
and hermitian models coincide.

\newsection{Main definitions}
The iterative procedure we are going to describe is based on three main
ingredients, the loop insertion operator, the first in the chain of loop
equations of our model and a suitable change of variables. It allows us
to calculate the genus $g$ contribution to the $n$-loop correlator for
(in principle) any $g$ and any $n$ and (also in practise) for any
potential. The possibility of going to an arbitrarily high genus is
provided by the loop equation while the possibility of considering an
arbitrarily high number of loops is provided by the loop insertion
operator. The change of variables allows us to consider an arbitrary
potential.

\subsection{Loop insertion operator}

The hermitian one matrix model is defined by the partition function
\beq
Z[g,N]=e^{N^2 F}=\int_{N\times N} d\phi\exp(-N\tr V( \phi)) \label{partition}
\eeq
where the integration is over hermitian $N\times N$ matrices and
\beq
V(\phi)=\sum_{j=1}^{\infty}\frac{g_j}{j}\,\:\phi^j
\label{potential}
\eeq
Expectation values (or averages) are defined in the usual way  as
\be
\LA  Q(\phi)\RA = \frac 1Z \int d\phi\exp(-N\tr V( \phi)) \, Q(\phi)\;.
\label{averages}
\ee
We introduce the generating functional (the $1$-loop average)
\beq
W(p)=\frac{1}{N}\sum_{k=0}^{\infty}\langle \tr\phi^{k}\rangle
/p^{k+1}
\eeq
and the $s$--loop correlator $(s\geq2)$
\beq
W(p_1,\ldots,p_s)=N^{s-2}\sum_{k_1,\ldots,k_s=1}^{\infty}
\langle \tr\phi^{k_1} \ldots
\tr\phi^{k_s}\rangle_{conn.}
/p_1^{k_1+1}\ldots p_s^{k_s+1}
\eeq
where $conn.$ refers to the connected part. One can rewrite the last two
equations as follows
\be
W(p_1,\ldots,p_s)=N^{s-2} \LA \tr\frac{1}{p_1-\phi} \ldots
\tr\frac{1}{p_s-\phi} \RA_{conn.}\,.
\ee
As is explained in Ref.~\cite{Kaz89}, these quantities are associated
with the (Laplace images of the) sum over discretized open
surfaces with $s$ boundaries.

The correlators can be obtained from the free energy, $F$,
by application of the loop insertion operator,
$\frac{d}{dV(p)}$:
\beq
W(p_1,\ldots, p_s)=\frac{d}{dV(p_s)}\;\frac{d}{dV(p_{s-1})}\ldots
\frac{dF}{dV(p_1)}
\label{loopaverage}
\eeq
where
\beq
\frac{d}{dV(p)}\equiv
-\sum_{j=1}^{\infty}\frac{j}{p^{j+1}}\frac{d}{dg_{j}}\;.
\label{2.8}
\eeq
Notice that \eq{loopaverage} can be rewritten as
\beq
W(p_1,\ldots, p_s)=\frac{d}{dV(p_s)}\;\frac{d}{dV(p_{s-1})}\ldots
\frac{d}{dV(p_2)} W(p_1)
\label{loopgener}
\eeq
which  shows that if the $1$-loop correlator is known for an arbitrary
potential, all multi-loop correlators can be calculated.
This is why it is named above as the generating functional.

With the normalization chosen above, the genus expansion of the
correlators reads
\beq
W(p_1,\ldots,p_s)=\sum_{g=0}^{\infty}\frac{1}{N^{2g}}\;
W_{g}(p_1,\ldots,p_s) \hspace{1.0cm} (s\geq 1)\,. \label{genus}
\eeq
Similarly we have
\beq
F=\sum_{g=0}^{\infty}\frac{1}{N^{2g}}F_g
\label{freeen}
\eeq
for the genus expansion of the free energy.

\subsection{The loop equation}

The first in the chain of loop equations for \hm\ can conveniently be written
as~\cite{Mak91}
\beq
\ci W(\om) = (W(p))^2 +\frac{1}{N^2} W(p,p)
\label{loop}
\eeq
where $V(\om)=\sum_{j}g_j\om^{j}/j $ and
$C$ is a curve which encloses the singularities of $W(\omega)$ and
not the point $\om=p$. This contour integration acts as a projector
picking up the coefficients of $p^{-1}$. Due to
\eq{loopgener} the second term on the
right hand side of the loop equation \rf{loop} is expressed via $W(p)$,
so that~\rf{loop} is a closed equation which determines  this quantity.

To leading order in $1/N^2$ one can ignore the last term
in~\eq{loop} and assuming that the singularities of $W(\om)$ consist of
only one cut on the real axis and that $W(p)$ behaves as $1/p$ as
$p\rightarrow \infty$ one finds~\cite{Mig83}
\beq
W_{0}(p)=\oh \ci \left\{\frac{(p-x)(p-y)}{(\om-x)(\om-y)}\right\}^{1/2}
\label{zero}
\eeq
where $x$ and $y$ are determined by the matrix model potential in the
following way
\beq
\cim\; \frac{V'(\om)}{\sqrt{(\om-x)(\om-y)}}=0\,,
\hspace{1.0cm}
\cim\; \frac{\om V'(\om)}{\sqrt{(\om-x)(\om-y)}}=2\,.
 \label{xandy}
\eeq
This genus zero solution will be used below in the iterative
procedure of solving the loop equation.

Inserting the genus expansion~\rf{genus} in~\eq{loop} it appears that
$W_g(p)\,,\;g\geq1$, obeys the following equation
\beq
\left\{\hat{K}-2W_{0}(p)\right\}W_g(p)=\sum_{g'=1}^{g-1}
W_{g'}(p)\;W_{g-g'}(p)
+\frac{d}{d V(p)}W_{g-1}(p) \label{loopg}
\eeq
where  $\hat{K}$ is a linear operator, namely
\beq
\hat{K}f(p)\equiv \ci f(\om)\,.
\label{hatK}
\eeq
In~\eq{loopg} $W_g(p)$ is expressed entirely in terms of
$W_{g_i}(p),\;\;g_i<g$. This is what makes it possible to
develop the iterative procedure mentioned in the Introduction.

\subsection{The new variables}

To characterize the matrix model potential we introduce instead of the
couplings $g_j$ the moments $M_k$ and $J_k$ defined by
\bea
M_{k}&=&\cim\, \frac{V'(\om)}{(\om-x)^{k+1/2}\,(\om-y)^{1/2}}
\hspace{1.7cm} k\geq 1\,,
\label{moment1}\\
J_{k}&=&\cim\, \frac{V'(\om)}{(\om-x)^{1/2}\,(\om-y)^{k+1/2}}
   \hspace{1.7cm} k\geq 1 \,.\label{moment2}
\eea
These moments depend on the coupling constants $g_j$'s both explicitly
and via $x$ and $y$ which are determined by Eq.~\rf{xandy}:
\bea
M_k&=&g_k+\left(\half x +
(k+\half)y\right)g_{k+1}+\ldots\;,
\\ J_k&=&g_k+\left((k+\half)x +\half y\right)g_{k+1}+\ldots\;.
\eea
Notice that $M_k$ and $J_k$ depend explicitly only on $g_j$ with $j\geq k$.

There are several motivations for
introducing these new variables. First, as we shall
see below,
for each term in the genus expansion of the partition function and the
correlators, the dependence on the infinite set of coupling constants
will arrange into a simple function of a {\it finte\/} number of the
moments.
Moreover, these new variables reflect
more directly than
the coupling constants
the possible critical behaviour of the matrix model. Let
us briefly describe how this comes about.

Performing the contour integral in~\rf{zero} by
taking residuals at $\om=p$ and
$\om=\infty$ one finds
\beq
W_0(p) =\frac{1}{2}\left\{V'(p)-M(p)\sqrt{(p-x)(p-y)}\right\}
\label{zeroint}
\eeq
where $M(p)$ is a polynomial in $p$ of degree two less than $V(p)$.
As already discussed in the Introduction, $W_0(p)$ can also be determined
by an analysis of  the matrix
model in the eigenvalue picture~\cite{BIPZ78}. Requiring that $W_0(p)$
is analytic in
the complex plane except for a branch cut $[y,x]$ on the real axis and
behaves as $1/p$ for $p\rightarrow\infty$
corresponds to requiring that the eigenvalue density, $\rho (\lambda)$,
has support only on the interval $[y,x]$ and is normalized to 1.
The eigenvalue density is in this situation given by
\beq
\rho(\lambda) =\frac{1}{\pi} M(\lambda) \sqrt{(\lambda-y)(x-\lambda)}
\hspace{0.7cm} y\leq \lambda \leq x \,.\label{evden}
\eeq

This function vanishes under normal circumstances as a square root at
both ends of its support.
Critical behaviour arises when some of the roots of $M(\lambda)$
approach $x$ or $y$. For a
non-symmetric potential the so called  $m^{th}$ multi-critical point is
reached when $(m-1)$ extra zeros accumulate at either $x$ or $y$.
Comparing~\rf{zero} and~\rf{zeroint} it appears that
\beq
M_k \propto
{\left.\frac{d^{(k-1)}M(\lambda)}{d\lambda^{k-1}}\right|}_{\lambda=x}
{}~~~~~,~~~~~
J_k \propto
{\left. \frac{d^{(k-1)}M(\lambda)}{d\lambda^{k-1}}\right|}_{\lambda=y }
\eeq
so the condition for being at the $m^{th}$ multi-critical point is
simply $M_1=M_2=\ldots=M_{k-1}=0$ and $M_k \neq 0$
if the extra zeros accumulate at $x$ and
$J_1=J_2=\ldots =J_{k-1}=0$, $J_k \neq 0$ if the extra zeros accumulate at $y$.
For the symmetric potential ($V(\phi)=V(-\phi)$)
the eigenvalues will be distributed
symmetrically around zero and hence  confined to an
interval of the type $[-\sqrt{z},\sqrt{z}]$ in the one arc case.
 In a situation like this
the $m^{th}$ multi-critical point is characterized by the eigenvalue
density having $(m-1)$ extra zeros at both $-\sqrt{z}$ and $\sqrt{z}$.
Reexamining~\rf{moment1} and~\rf{moment2}
it appears that
for the symmetric potential
\beq
M_k=(-1)^{k+1}J_k\,, \label{momsym}
\hspace{0.7cm}(V(\phi)=V(-\phi))\,.
\eeq
Hence in this case we have only one set of moments and the condition for
being at the $m^{th}$ multi-critical point is the vanishing of the first
$(m-1)$ of these.

This
formalism obviously allows for a treatment of a more general
situation where $m$ extra zeros accumulate at $x$ and $n$ extra zeros
accumulate at $y$. Such multi-critical models have been studied
in Ref.~\cite{Morrismul}. We will however
restrict ourselves to the traditional models.

The superiority of the moments defined in~\rf{moment1} and~\rf{moment2}
as compared to the coupling constants will become
even more clear when we consider the double scaling limit in
Section~\ref{dsl}.
By then it will also become evident why these new moments are more
convenient than those
originally introduced in Ref.~\cite{ACM92}.

\newsection{The iterative procedure}

Our iterative solution of the loop equation results in a certain
representation of the free energy and the correlators in terms of the moments.
In this section we describe the structure of $W_g(p)$ and $F_g$
and show that the
iterative procedure  can be conveniently
formulated by referring to the eigenvectors of a linear operator, $\hat{K}$.
We prove that
the genus $g$ contribution to the $s$-loop correlator depends at most on
 $2\times(3g-2+s)$ lower moments ($2\times(3g-2)$ for the partition
function)
 for $g >0$.
We perform explicit calculations up to genus two.

\subsection{The structure of $F_g$ and $W_g(p)$ \label{struct} }

Our iterative procedure of solving the loop equation results in the
following representation of the genus $g$ contribution to the free energy
\beq
F_g=\sum_{\a_j>1,\atop \b_i>1}\langle \a_1 \ldots \a_s ; \b_1 \ldots
\b_l |\a, \b,\g \rangle_g
{M_{\a_1}\ldots M_{\a_s}J_{\b_1}\ldots J_{\b_l}\over M_1^{\a}J_1^{\b}d^{\g}}
\hspace{0.5cm} g \geq1\;,
\label{fgen}
\eeq
where $d=x-y$ is the distance between the endpoints of the support of
the eigenvalue distribution, the brackets denote rational
numbers and $\a$, $\b$ and $\g$ are non-negative integers. The indices
$\a_1,\ldots,\a_s,\b_1,\ldots,\b_l$ take values in the interval
$[2,3g-2]$ and the summation is over sets of indices.
 In particular $F_g$ depends on at most $2\times (3g-2)$
moments. Furthermore, since nothing allows us to distinguish between
$x$ and $y$, $F_g$ must be invariant under the interchange of the two.
Hence one gets
\beq
F:\hspace{0.5cm}
\langle \a_1 \ldots \a_s ; \b_1 \ldots \b_l |\a, \b, \g
\rangle = (-1)^{\g}
\langle \b_1 \ldots \b_l ; \a_1 \ldots \a_s |\b, \a, \g \rangle\;.
\label{bracket}
\eeq

There are certain restrictions on the integers which enter \eq{fgen}.
Let us denote by $N_M$ and $N_J$ the total powers of $M$'s and $J$'s,
respectively, {\it i.e.}
\beq
N_M=s-\a \,,\hspace{0.7cm} N_J=l-\b\,.
\eeq
Then it holds that
\beq
N_M \leq 0,\hspace{1.0cm} N_J \leq 0
\label{number0}
\eeq
and
\bea
F_g:& \hspace{0.5cm} N_M+N_J = 2-2g\,,&  \label{inv1}\\
F_g:&\hspace{0.5cm} \sum_{i=1}^{s}(\a_i-1)+\sum_{j=1}^{l}(\b_j-1)
+\g=4g-4\,. & \label{inv2F}
\eea
The relation~\rf{inv1} follows from the fact that the partition function
$Z=e^{\sum_gN^{2-2g}F_g}$ is invariant under simultaneous rescalings of
$N$ and the eigenvalue density, $\rho(\lambda)$; \mbox{$N\rightarrow k\cdot
N$},
\mbox{$\rho(\lambda)\rightarrow \frac{1}{k}\rho(\lambda)$}. The
relation~\rf{inv2F} follows from the invariance of $Z$ under rescalings
of the type \mbox{$N\rightarrow k^2\cdot N$}, \mbox{$g_j\rightarrow
k^{j-2}g_j$}.
Finally the following inequality must be fulfilled:
\beq
F_g:\hspace{0.5cm} \sum_{i=1}^{s}(\a_i-1)+\sum_{j=1}^{l}(\b_j-1)
\leq 3g-3 \label{homF}
\eeq
This requirement becomes more transparent when we consider the double
scaling limit in Section~\ref{dsl}. In combination with
Eq.~\rf{inv2F} it gives
\beq
g-1\leq \g \leq 4g-4\,.
\label{boundg}
\eeq

To explain the structure of $W_g(p)$, let
us introduce the basis vectors $\chi^{(n)}(p)$ and
$\Psi^{(n)}(p)$:
\bea
\chi^{(n)}(p)&=&\frac{1}{M_1}\left\{\Phi_x^{(n)}(p)
         -\sum_{k=1}^{n-1}\chi^{(k)}(p)M_{n-k+1}
          \right\} \hspace{1.0cm}n\geq 1\;,
\label{chi}\\
\Psi^{(n)}(p)&=&\frac{1}{J_1}\;\left\{\Phi_y^{(n)}(p)
         -\sum_{k=1}^{n-1}\Psi^{(k)}(p)J_{n-k+1}
          \right\}\hspace{1.15cm} n\geq 1
\label{Psi}
\eea
where
\bea
\Phi_x^{(n)}(p)&=&(p-x)^{-n}\,\left\{(p-x)(p-y)\right\}^{-1/2}
\hspace{1.0cm} n \geq 0\;,\label{phix}\\
\Phi_y^{(n)}(p)&=&(p-y)^{-n}\,\left\{(p-x)(p-y)\right\}^{-1/2}
\hspace{1.0cm} n \geq 0\;. \label{phiy}
\eea
It is easy to show for the operator $\hat{K}$ defined by \eq{hatK} that
\bea
\left\{\hat{K}-2W_0(p)\right\}\chi^{(n)}(p) &= & \frac{1}{(p-x)^n}
\hspace{1.0cm} n \geq 1\;,
\label{Kchi}\\
\left\{\hat{K}-2W_0(p)\right\}\Psi^{(n)}(p) &=&\frac{1}{(p-y)^n}
\hspace{1.0cm} n\geq 1
\label{KPsi}
\eea
and that the kernel of $\left\{\hat{K}-2W_0(p)\right\}$ is spanned by
$\Phi_x^{(0)}(p)=\Phi_y^{(0)}(p)$.

Since $W_g(p)$
can be obtained from $F_g$ according to \eq{loopaverage},
the representation \rf{fgen} implies%
\footnote{ A similar formula is proven in Ref.~\cite{ACM92} for a different
definition of the moments and of the basis vectors.}
\beq
W_g(p)=\sum_{n=1}^{3g-1}\left\{A_g^{(n)}\chi^{(n)}(p)
       + D_g^{(n)} \Psi^{(n)}(p)\right\}\hspace{0.5cm}g\geq 1\;.
\label{conjecture}
\eeq
We do not add any multiple of $\Phi_x^{(0)}(p)$ or $\Phi_y^{(0)}(p)$.
Doing so would contradict the boundary condition $W(p)\rightarrow 1/p$
for $p\rightarrow \infty$ since this behaviour was already obtained for
genus zero. We note that this structure of $W_g(p)$ is in agreement with
the assumption~\cite{Dav90} that $W_g(p)$ is analytic in the complex plane
except for a branch cut $[y,x]$ on the real axis.

The coefficients $A_g^{(n)}$ are of the same structure as
$F_g$ and the
relation~\rf{number0} still holds. However in this case the indices
$\a_1,\ldots,\a_s,\b_1,\ldots,\b_l$ take values in the
interval $[2,3g-n]$. Hence $W_g(p)$ depends on at most $2\times (3g-1)$
moments. The invariance of the partition function under the
rescalings described above has the following implications for the
structure of $A_g^{(n)}$:
\bea
A_g^{(n)}:&\hspace{0.5cm} N_M+N_J = 2-2g\;,& \label{numberA}\\
A_g^{(n)}:&\hspace{0.5cm} \sum_{i=1}^{s}(\a_i-1)+\sum_{j=1}^{l}(\b_j-1)
+\g=4g-2-n\;.& \label{inv2A}
\eea
We also have an analogue of~\rf{homF} for $A_g^{(n)}$. It reads
\beq
A_g^{(n)}:\hspace{0.5cm} \sum_{i=1}^{k}(\a_i-1)+\sum_{j=1}^{l}(\b_j-1)
\leq 3g-n-1 \,. \label{homA}
\eeq
Again we refer to Section~\ref{dsl} for further explanations.
However, we note that combining~\rf{inv2A} and~\rf{homA} one gets again
the bound~\rf{boundg} on $\g$.

As was the
case for $F_g$, $W_g(p)$ must be invariant under the interchange of
$x$ and $y$. This means that
$D_g^{(n)}$ must appear from $A_g^{(n)}$ by the replacements
$d\rightarrow -d$, $J\leftrightarrow M$. (We note that we do not have a
relation like~\rf{bracket} for the $A_g^{(n)}$'s.)
Furthermore, $W_g(p)$ must be an odd
function of $p$ for a symmetric potential.
This implies that
$A_g^{(n)}=(-1)^n D_g^{(n)}$.

That the structure of $F_g$ and $W_g(p)$ actually is as described in this
section can be proven by induction.
We will not go through the proof here. Instead we refer to
Refs.~\cite{ACM92,complex}.
In the latter reference a formula somewhat similar
to~\rf{conjecture} was proven for the complex matrix
model. However the strategy of the proof will be evident from
Sections~\ref{Wg} and~\ref{Fg} where we describe the iterative procedures
which allow us to
calculate $W_g(p)$ and $F_g$ for any $g$ starting from $W_0(p)$.

\subsection{The iterative procedure for determining $W_g(p)$
\label{Wg} }

According to~\eq{loop}, we need to calculate $W_0(p,p)$ in order to
 start the iterative procedure.
To do this we write the loop insertion operator as
\beq
\frac{d}{dV(p)}=\frac{\partial}{\partial V(p)}+
                \frac{dx}{dV(p)}\frac{\partial}{\partial x}
              +\frac{dy}{dV(p)}\frac{\partial}{\partial y}
\eeq
where
\beq
\frac{\partial}{\partial V(p)}=-\sum_{j=1}^{\infty}\frac{j}{p^{j+1}}
                             \frac{\partial}{\partial g_j}\,.
\eeq
The derivatives $dx/dV(p)$ and $dy/dV(p)$ can be obtained
from~\rf{xandy} and read
\beq
\frac{dx}{dV(p)} = \frac{1}{M_1} \Phi_x^{(1)},
\hspace{1.0cm}
\frac{dy}{dV(p)} = \frac{1}{J_1} \Phi_y^{(1)}\,.
\label{xandyderiv}
\eeq
Using the relation
\beq
\frac{\partial}{\partial V(p)}V'(\om) =
\frac{\partial}{\partial p}\, \frac{1}{p-\om}\,,
\label{trick}
\eeq
one finds~\cite{jjm}
\beq
W_0(p,p)=\frac{(x-y)^2}{16(p-x)^2(p-y)^2}\,.
\label{twoloop}
\eeq
This enables us to find $W_1(p)$ and we see that it is of the form
\rf{conjecture} with
\bea
A_1^{(1)}=-\frac{1}{8d},&\hspace{0.8cm}& A_1^{(2)} = \frac{1}{16} ,
\label{genus1a}\\
D_1^{(1)}=\frac{1}{8d},&\hspace{0.8cm}& D_1^{(2)} = \frac{1}{16}.
\label{genus1b}
\eea
Carrying on the iteration process is straightforward. In each step one
must calculate the right hand side of the loop equation~\rf{loop}.
Decomposing the result obtained into fractions of the type
$(p-x)^{-n}$, $(p-y)^{-n}$ allows one to identify immediately the
coefficients $A_g^{(n)}$ and $D_g^{(n)}$.

To calculate $W_g(p,p)$ it is
convenient to write the loop insertion operator as
\beq
\frac{d}{dV(p)} = \sum_{n}\frac{dM_n}{dV(p)}\frac{\partial}{\partial M_n}
+\sum_{j}\frac{dJ_j}{dV(p)}\frac{\partial}{\partial J_j}
+\frac{dx}{dV(p)} \frac{\partial}{\partial x}
+\frac{dy}{dV(p)} \frac{\partial}{\partial y}
\label{loopins}
\eeq
where
\bea
\frac{dM_n}{dV(p)}&=&-\frac{1}{2}(p-x)^{-n-1/2}\,(p-y)^{-3/2}
                   -(n+1/2)\Phi_x^{(n+1)}(p)\nonumber \\
&&      +\frac{1}{2}\left\{
      \sum_{i=1}^{n}(-1)^{n-i}M_i\left(\frac{1}{x-y}\right)^{n-i+1} \,
                         +(-1)^n J_1\left(\frac{1}{x-y}\right)^n
                    \right\} \frac{dy}{dV(p)}  \nonumber \\
&&                   +(n+1/2)M_{n+1}\frac{dx}{dV(p)}\,.
\label{dMdV}
\eea
The derivatives $dx/dV(p)$ and $dy/dV(p)$ are given by~\rf{xandyderiv}
and the function $\Phi_x^{(n)}$ was defined
in~\rf{phix}. Of course $dJ_n/dV(p)$ just appears from
$dM_n/dV(p)$ by the replacements $y\leftrightarrow x$ and
$J\leftrightarrow M$.
We note that there is no simplification of the algorithm in the case of
the symmetric potential. We can only put $x=-y=\sqrt{z}$ at the end of the
calculation.
This complication stems of course from the fact
that we have to keep the odd coupling constants
in the loop insertion operator
until all differentiations have been performed. Only hereafter they can
be put equal to zero. The same is not true in the double scaling limit
however. We will come back to this later.

By taking a closer look at the loop
insertion operator~\rf{loopins} and bearing in mind the
results~\rf{twoloop}, \rf{genus1a} and~\rf{genus1b}, it is easy to
convince oneself that $A_g^{(n)}$ and $D_g^{(n)}$ depend only on $x$ and
$y$ via $(x-y)$ and have the structure shown in
equation~\rf{fgen}.
Furthermore it appears that $W_g(p_1,\ldots,p_s)$
depends on at most $2\times (3g-2+s)$ parameters.
The results for $g=2$ obtained with the aid of
{\sl Mathematica} read
\begin{eqnarray}
A_2^{(1)}& = &
\frac{201\,}{256\,{{d}^5}\,{{J_{1}}^2}}
 - \frac{67\,J_{2}}{128\,{{d}^4}\,{{J_{1}}^3}}
-\frac{5\,J_{3}}{32\,{{d}^3}\,{{J_{1}}^3}}
 +\frac{49\,{{J_{2}}^2}}{256\,{{d}^3}\,{{J_{1}}^4}}
      \nonumber \\*
&& +
 \frac{57}{64\,d^5\,J_{1}\,M_{1}}
 -\frac{11\,J_{2} }{128\,d^4\,{J_{1}}^2\,M_{1}}
      +
   \frac{49\,{M_{2}}^2}
     {256\,{d}^3\,{M_{1}}^4} \nonumber \\*
&& +
   \frac{201\,}{256\,{{d}^5}\,{M_{1}}^2}
 + \frac{22\,M_{2}}{128\,{{d}^4}\,{{J_{1}}}\,{{M_{1}}^2}}
 - \frac{J_{2}\,M_{2} }{64\,{{d}^3}\,{J_{1}}^2\,{M_{1}}^2}
      \nonumber \\*
&&+
   \frac{67\,M_{2}}{128\,{d}^4\,{M_{1}}^3}
-\frac{5\,M_{3}}{32\,{d}^3\,{M_{1}}^3}\;,
 \nonumber \\
A_2^{(2)}&= &
-\frac{57}{128\,d^4\,J_{1}\,M_{1}}
+ \frac{8\,J_{2}}{128\,d^3\,{J_{1}}^2\,M_{1}}
      -    \frac{49\,{M_{2}}^2}
     {256\,{d}^2\,{M_{1}}^4} \nonumber \\*
&&
  - \frac{201}{256\,{{d}^4}\,{M_{1}}^2}
 - \frac{3\,M_{2}}{128\,{{d}^3}\,{{J_{1}}}\,{{M_{1}}^2}} +
       \frac{J_{2}\,M_{2} }{128\,{{d}^2}\,{{J_{1}}^2}\,{{M_{1}}^2}}
      \nonumber \\*
&&
  - \frac{67\,M_{2}}{128\,{{d}^3}\,{M_{1}}^3}
  +\frac{5\,M_{3}}{32\,{{d}^3}\,{M_{1}}^3}\;,
     \nonumber \\
A_2^{(3)} &= &
 \frac{49\,{M_{2}}^2}{256\,d \,{M_{1}}^4}
 -\frac{5\,M_{3}}{32\,d\,{M_{1}}^3}
+\frac{67\,M_{2}}{128\,{d}^2\,{M_{1}}^3}
\nonumber \\*
&&
+\frac{201}{256\,{{d}^3}\,{{M_{1}}^2}} +
   \frac{15}{128\,{d}^3\,{J_{1}}\,M_{1}}-
\frac{5\,J_{2}}{128\,d^2\,{J_{1}}^2\,M_{1}}\;,
     \nonumber \\
A_2^{(4)} &=&
   - {{49\,M_{2}}\over {128\,d \,{{M_{1}}^3}}}
-{{189}\over {256\,{{d}^2}\,{{M_{1}}^2}}}\;, \nonumber \\
A_2^{(5)} &=&
{{105}\over {256\,d \,{{M_{1}}^2}}}\;;\nonumber \\
&& \nonumber\\
D_2^{(1)} &= & A_2^{(1)}\, \left( M\longleftrightarrow J,\,\,
d\longrightarrow -d\right)\,,\nonumber \\
D_2^{(2)} & = &A_2^{(2)}\, \left( M\longleftrightarrow J,\,\,
d\longrightarrow -d\right)\,,
 \nonumber \\
D_2^{(3)} &= &A_2^{(3)}\, \left( M\longleftrightarrow J,\,\,
d\longrightarrow -d\right)\,,\nonumber \\
D_2^{(4)} & = &A_2^{(4)}\, \left( M\longleftrightarrow J,\,\,
d\longrightarrow -d\right)\,,
     \nonumber \\
D_2^{(5)} &= &A_2^{(5)}\, \left( M\longleftrightarrow J,\,\,
d\longrightarrow -d\right)\,.
\end{eqnarray}
The genus two contribution to $W(p)$ is now determined by \eq{conjecture}.

\subsection{The iterative procedure for $F_g$ \label{Fg} }

In this section we present an algorithm which allows one to determine
$F_g$, as soon as the result for $W_g(p)$ is known.
The strategy consists
in writing the basis vectors $\chi^{(n)}(p)$ and $\Psi^{(n)}(p)$ as
derivatives with respect to $V(p)$. It is easy to verify that the
following relations hold
\bea
\chi^{(1)}(p)&=& \frac{dx}{dV(p)}\,, \label{xdV}\\
\Psi^{(1)}(p)&=& \frac{dy}{dV(p)}\,, \\
\chi^{(2)}(p)&=& \frac{d}{dV(p)}\left\{-\frac{2}{3}\ln M_1
-\frac{1}{3}\ln d\right\}\,, \\
\Psi^{(2)}(p)&=&\frac{d}{dV(p)}\left\{-\frac{2}{3}\ln J_1
-\frac{1}{3}\ln d\right\}\,.
\eea
Combining this with the results~\rf{genus1a} and~\rf{genus1b} one
immediately finds
\beq
F_1=-\frac{1}{24}\ln M_1 -\frac{1}{24}\ln J_1 -\frac{1}{6} \ln d
\label{hermF1}
\eeq
which coincides with the expression of Ref.~\cite{ACM92}.

In the general case things are not quite as simple. The basis vectors
can not be written as total derivatives. This is of course in
accordance with the fact that the $A$ and $D$ coefficients now have a
more complicated dependence on the potential (cf.\ Section~\ref{Wg} ). However,
a rewriting of the basis vectors allows one to identify relatively
simply $W_g(p)$ as a total derivative. In the case of $\chi^{(n)}(p)$
this rewriting reads
\bea
\chi^{(n)}(p) &=&\frac{1}{M_1}\left\{
-\frac{1}{2n-1}\sum_{i=1}^{n-1}
(-1)^{n-i-1}\left\{\Phi_x^{(i)}-M_i\frac{dy}{dV(p)}\right\}
\left(\frac{1}{x-y}\right)^{n-i}\, \right.
\nonumber\\*
&& \left.
-\frac{2}{2n-1}\frac{dM_{n-1}}{dV(p)}
-\sum_{k=2}^{n-1}\chi^{(k)}M_{n-k+1}
\right\},
\hspace{0.7cm}n\geq 2
\eea
where $\Phi_x^{(n)}$ should be written as
\bea
\Phi_x^{(n)} &= &
 -\frac{1}{2n-1}\sum_{i=1}^{n-1}
(-1)^{n-i-1}\left\{\Phi_x^{(i)}-M_i\frac{dy}{dV(p)}\right\}
\left(\frac{1}{x-y}\right)^{n-i}
\nonumber \\*
&&+M_n \frac{dx}{dV(p)} - \frac{2}{2n-1}
\frac{dM_{n-1}}{dV(p)}
   \hspace{0.5cm}
 n\geq2 \,,\\
\Phi_x^{(1)}&=&M_1 \frac{dx}{dV(p)}\,.
\eea
The basis vector $\chi^{(1)}(p)$ should of course still be written as
in~\rf{xdV}. The rewriting of the $\Psi^{(n)}$'s is analogous to that
of the $\chi^{(n)}$'s. It can be obtained by performing the replacements
$J\leftrightarrow M$ and $x\leftrightarrow y$ in the formulas above.

 By means of these rewritings we have been able to determine $F_2$. The
result reads
\bea
F_2&=&
  -{{119}\over {7680\,{{J_1}^2}\,{{d }^4}}} -
   {{119}\over {7680\,{{M_1}^2}\,{{d }^4}}}
 +    {{181\,J_2}\over
     {480\,{{J_1}^3}\,{{d }^3}}}
-    {{181\,M_2}\over
     {480\,{{M_1}^3}\,{{d }^3}}}\nonumber \\*
&&+
   {{3\,J_2}\over
     {64\,{{J_1}^2}\,M_1\,{{d }^3}}}
 -    {{3\,M_2}\over
     {64\,J_1\,{{M_1}^2}\,{{d }^3}}} -
   {{11\,{{J_2}^2}}\over
     {40\,{{J_1}^4}\,{{d }^2}}}
-{{11\,{{M_2}^2}}\over
     {40\,{{M_1}^4}\,{{d }^2}}}\nonumber \\*
&&  +
   {{43\,M_3}\over
     {192\,{{M_1}^3}\,{{d }^2}}}
+   {{43\,J_3}\over
     {192\,{{J_1}^3}\,{{d }^2}}} +
   {{J_2\,M_2}\over
     {64\,{{J_1}^2}\,{{M_1}^2}\,{{d }^2}}}
-    {{17}\over {128\,J_1\,M_1\,{{d }^4}}}  \nonumber \\*
&&+
   {{21\,{{J_2}^3}}\over
     {160\,{{J_1}^5}\,d }} -
   {{29\,J_2\,J_3}\over
     {128\,{{J_1}^4}\,d }} +
   {{35\,J_4}\over {384\,{{J_1}^3}\,d }} -
   {{21\,{{M_2}^3}}\over
     {160\,{{M_1}^5}\,d }} \
\nonumber \\*
&&+
   {{29\,M_2\,M_3}\over
     {128\,{{M_1}^4}\,d }} -
   {{35\,M_4}\over {384\,{{M_1}^3}\,d }}\,.
\label{f2complete}
\eea
It is obvious from the formulas above that $F_g$ depends
for a non-symmetric potential on at most $2\times (3g-2)$ moments.
Furthermore,
for a symmetric potential, $F_g$ is a sum of two identical terms and
depends on only at most $3g-2$ moments.
A consequence of this doubling for \dsl\ will be discussed
in Section~\ref{F_g}.

\newsection{The double scaling limit}

It is easy to determine which terms in the explicit expressions for
$F_g$ and $W_g(p)$ determined in the previous section that contribute in
the double scaling limit. However it is rather time consuming to
determine $F_g$  and $W_g(p)$ away from the double scaling limit.
In this section we develop an algorithm which gives us directly the
result in the double scaling limit.
Using this algorithm we
calculate the correlators and the partition function
explicitly up to genus four and describe their general structure.

\subsection{Multi-critical points \label{dsl} }

Let us consider first the case of the symmetric potential. We hence have
$x=-y=\sqrt{z}$ and $J_k=(-1)^{k+1}M_k$ for all values of $k$. As is mentioned
earlier, the $m^{th}$ multi-critical point is characterized by the
eigenvalue density having $(m-1)$ extra zeros accumulating at both
$-\sqrt{z}$ and $\sqrt{z}$, and the condition for being at this point
is the vanishing of the first $(m-1)$ moments. For the $m^{th}$
multi-critical model the double scaling limit of the correlators is obtained
by fixing the ratio of any given coupling and, say $g_2$, to its critical
value and setting
\bea
p^2&=&z_c+a\pi\,, \label{scalps} \\
z&=&z_c-a\Lambda^{1/m}\,.\label{scalzs}
\eea
The moments then scale as
\beq
J_k\sim M_k\sim a^{m-k},\hspace{1.0cm} k\in[1,m-1]\;.
\label{scalmom}
\eeq

Furthermore, it is well known that the genus $g$ ($g\geq 1$)
contribution to the
free energy has the following scaling behaviour
\beq
F_g \sim a^{(2-2g)(m+1/2)}\,. \label{Fscal}
\eeq
Bearing in mind that the structure of $F_g$ is as shown in
\eq{fgen}, one finds that the following relation must hold
\beq
\sum_{j=1}^s(m-\a_j)+\sum_{i=1}^l(m-\b_i)-(\a+\b)(m-1)
\geq m(2-2g)-g-1\,.
\eeq
Since the free energy should look the same for all multi-critical models,
we have
\beq
N_J+N_M\geq 2-2g \,, \label{Frel1}
\eeq
\beq
\sum_{j=1}^s(\a_j-1)+\sum_{i=1}^l(\b_i-1)\leq 3g-3\,.
\label{Frel2}
\eeq
We already know from the analysis in Section~\ref{struct} that the
equality sign must hold in~\rf{Frel1}. Only terms for which
$max(s,l) \leq m$ and for which
the equality
sign holds also in~\rf{Frel2} will contribute in the double scaling
limit. From \eq{inv2F} it follows that these terms will have
\be
\g = g-1.
\label{g-1}
\ee

A similar analysis can be carried out for the generating functional.
Here it is known that the genus $g$ contribution to $W(p)$ has the
following scaling behaviour
\beq
W_g(\pi,\Lambda)\sim a^{(1-2g)(m+1/2)-1}
\label{scal}
\eeq
with the exception of $W_0(\pi,\Lambda)$ which also contains a
non-scaling part. From~\rf{chi} and~\rf{Psi} it appears that the basis
vectors scale as
\beq
\chi^{(n)}\sim \Psi^{(n)} \sim a^{-m-n+1/2}\,.
\eeq
Remembering that the structure of the $A$ and $D$ coefficients is as
shown in~\rf{fgen}, one finds that the following relation must hold for
both $A_g^{(n)}$ and $D_g^{(n)}$
\beq
\sum_{j=1}^s(m-\a_j)+\sum_{i=1}^l(m-\b_i)
-(\a+\b)(m-1) \geq m(2-2g)+n-g-1\,.
\eeq
again provided $max(s,l)\leq m$.
Since also the generating functional should look the same for all
multi-critical models, it follows that $A_g^{(n)}$ and $D_g^{(n)}$ must
satisfy the following conditions
\beq
N_J+N_M\geq 2-2g \,,\label{NA}
\eeq
\beq
\sum_{j=1}^s(\a_j-1)+\sum_{i=1}^l(\b_i-1)\leq 3g-n+1 \,.
\label{homo}
\eeq
Here we recognize the homogeneity requirement~\rf{homA}.
We know from Section~\ref{struct} that the equality sign always holds
in~\rf{NA}.
Only terms for which $max(s,l)\leq m$ and for which
the equality sign holds also in~\rf{homo}
 will contribute in the double scaling limit. We note this means
that $\g=g-1$ for these terms  (cf.\ \eq{inv2A}).

Let us turn now to the case of the non-symmetric potential, and let us
consider the $m^{th}$ multi-critical point assuming that the extra zeros
accumulate at $x$. To obtain the double scaling limit of the correlators,
we
first fix as before the ratio between any given coupling and the
first one to its critical value. Then we scale $p$ and $x$ in the
following way keeping however $y$ fixed:
\bea
p&=&x_c+a\pi\,, \label{scalp}\\
x&=&x_c-a\Lambda^{1/m}\,.\label{scalx}
\eea
Under these circumstances the $J$-moments do not scale but $M_k$ behaves
as
\beq
M_k\sim a^{m-k} \hspace{0.7cm} k\in[1,m-1]
\eeq
Using again~\rf{Fscal} we find that the only terms which survive the
double scaling limit for $F_g$ are those for which
\beq
 N_M= 2-2g \,,\label{NMas}
\eeq
\beq
\sum_{j=1}^s(\a_j-1) = 3g-3\,. \label{homas}
\eeq
Comparing \eq{NMas} with \eq{inv1}, we see that
all the $J$-dependence disappears when
the prescription for taking the double scaling limit is as in~\rf{scalp}
and~\rf{scalx}. Furthermore,
comparing~\rf{homas} with~\rf{inv2F} we find that
all remaining terms have $\g=g-1$ as in the symmetric case.
The relations~\rf{NMas} and~\rf{homas} will be
exploited in Section~\ref{lkontsevich}.

Let us carry out the analysis of $W_g(p)$ also
for the non-symmetric case
since this will give us an additional information about the structure of
the $D$ and $A$ coefficients.
For the basis vectors we have
\beq
\chi^{(n)} \sim a^{-m-n+1/2}\,,\hspace{0.7cm} \Psi^{(n)}\sim a^{-1/2}\,.
\eeq
By exploiting again the known scaling behaviour~\rf{scal} of $W_g(p)$,
one can just as in the symmetric case derive certain homogeneity
conditions that the $A$ and $D$ coefficients must fulfill. One finds
\beq
A_g^{(n)}: \hspace{0.2cm} N_M\geq 2-2g,
\hspace{0.7cm}
 \sum_{j=1}^s(\a_j-1) \leq 3g-n-1 \,; \label{A}
\eeq
\beq
D_g^{(n)}: \hspace{0.2cm} N_M\geq 1-2g, \hspace{0.7cm}
 \sum_{j=1}^s(\a_j-1) \leq 3g-1\,. \label{D}
\eeq
The conditions are different for the $A$ and $D$ coefficients due to the
different scaling behaviour of the basis vectors. Terms that are
important in the double scaling limit are only those for which
$s\leq m$ and for which  the equality sign holds in both
relations in~\rf{A}
or both relations in~\rf{D}. Comparing the second relation in~\rf{D}
with~\rf{homo} we see that all $D$-terms vanish in the double scaling
limit. Furthermore, comparing~\rf{A} with~\rf{numberA}
we see that all $J$ dependent $A$ terms disappear.
Hence in the double scaling limit everything is expressed only in terms of
the $M$'s and $d$, and all terms have $\g=g-1$.

As is mentioned earlier, $D_g^{(n)}$ can be obtained from $A_g^{(n)}$ by
performing the replacements $x\leftrightarrow y$ and $J\leftrightarrow
M$ Hence we also have that the following homogeneity requirements should
be fulfilled
\beq
A_g^{(n)}: \hspace{0.2cm} N_J\geq 1-2g,
\hspace{0.7cm}
 \sum_{j=1}^l(\b_j-1) \leq 3g-1 \,;\label{A2}
\eeq
\beq
D_g^{(n)}: \hspace{0.2cm} N_J\geq 2-2g,\hspace{0.7cm}
 \sum_{j=1}^l(\b_j-1) \leq 3g-n-1 \,.\label{D2}
\eeq
These relations could of course be derived by analyzing the
scaling behaviour assuming that the extra zeros accumulate at $y$.
This implies fixing $x$ and replacing~\rf{scalp}
and~\rf{scalx} with
\bea
p&=&y_c-a\pi \,,\\
y&=&y_c+a\Lambda^{1/m}\,.
\eea
In
this situation only terms for which $l\leq m$ and for which
the equality sign holds in both
relations in~\rf{A2}
or both relations in~\rf{D2} will contribute in
the double scaling limit. For symmetry reasons, now all $A$ terms plus
$M$-dependent $D$ terms disappear.
Also the free energy is expressed under these circumstances entirely in
terms of the $J$'s and $d$.
We note that independently of the details of the prescription for taking
the double scaling limit, in this limit the multi-loop correlator
$W_g(p_1,\ldots,p_s)$ depends on at most $3g-2+s$ moments and the
free energy $F_g$ depends on at most $3g-2$ moments.

\subsection{Determining $W_g(p)$ in the d.s.l.}

The homogeneity conditions derived in the previous sections
 allow us to determine
whether a given term contributes to the double
scaling limit or not. As an example, let us consider the case of a
non-symmetric potential where the critical behaviour is associated with
the endpoint $x$. Using~\rf{A} we see that  only $A_2^{(5)}$,
the first term of $A_2^{(4)}$ and the first two terms of $A_2^{(3)}$
of the long list of complicated expressions for the $D$ and $A$
coefficients of $W_2(p)$ given in Section~\ref{Wg} survive in the double
scaling limit. Obviously it would be convenient to
have an algorithm which gives as output only terms which are relevant
for the continuum limit. An algorithm which gives us all potentially
relevant terms can be obtained by a slight
modification of the one described in Section~\ref{Wg}.
By potentially relevant we mean relevant for $m$ multi-critical models
with $m$ sufficiently large.

It is easy to
convince oneself that terms which are relevant for the double scaling
limit of $W_g(p)$ can only appear from terms which are relevant for the
double scaling limit of the right hand side of the loop equation.
Let us assume that we know the scaling relevant versions of
$A_G^{(n)}$ and $D_G^{(n)}$ for $G=1,\ldots,g$ and let us
assume that we want to calculate the double scaling limit of the right
hand side of the loop equation for $W_{g+1}$. First of all we note that
all terms in the basis vectors $\chi^{(n)}(p)$ and $\Psi^{(n)}(p)$ show
the same scaling behaviour so none of them can be ignored.
However we can replace all occurrences of $(p-y)$ with
$(x_c-y)=d_c$ in this limit. It is known from earlier
analysis~\cite{jm} that for
a $m$ multi-critical model $W_g(p,p)$ scales as
\beq
W_g(\pi,\pi)\sim a^{-2g(m+1/2)-2}. \label{scal1}
\eeq
Using equation~\rf{scal} it is easy to show that all the products in
the sum in~\rf{loop} have the same scaling behaviour. Hence, no
superfluous terms appear from the sum if we start from the double scaled
versions of the $W_G$, $G=1,\ldots g$.

 However, if we
calculate $W_g(p,p)$ by letting the loop insertion operator as written
in~\rf{loopins} act on $W_g(p)$, irrelevant terms will appear even if we
start from double scaled version of $W_g(p)$. By comparing~\rf{scal}
and~\rf{scal1}, we see that only operators in $d/dV(p)$ which lower the
power of $a$ by $m+3/2$ give rise to the relevant terms. Therefore, we should
discard all other operators. By examining carefully each term
in~\rf{loopins}, one finds that only the following part of the loop
insertion operator remains in the double scaling limit:
\beq
\frac{d}{dV(p)}_x=\sum_n\frac{dM_n}{dV(p)} \frac{\partial}{\partial M_n}
+\frac{dx}{dV(p)}\frac{\partial}{\partial x} \hspace{0.5cm} (d.s.l.)
\label{ddVx1}
\eeq
where
\bea
\frac{dM_n}{dV(p)}&=&-(n+1/2)\left\{\Phi_x^{(n+1)}(p)-
\frac{M_{n+1}}{M_1}\Phi_x^{(1)}(p)\right\}\hspace{0.5cm} (d.s.l.)\,,
\label{ddVx2}\\
\frac{dx}{dV(p)}&=&\frac{1}{M_1}\Phi_x^1(p)
\hspace{0.5cm} (d.s.l.) \label{ddVx3}
\eea
and
\beq
\Phi_x^{(n)}(p)= (p-x)^{-n}\left\{d_c\,(p-x)\right\}^{-1/2}
\label{ddVx4} \hspace{0.5cm} (d.s.l.).
\eeq

We are now in a position to calculate the double scaling limit of the
right hand side of the loop equation for $W_{g+1}$ provided we know the
double scaled versions of $W_1(p),\ldots, W_g(p)$. We see that, since all
the $y$-dependence has disappeared, we do not have to perform any
decomposition of the result. This is in accordance with the outcome of
the analysis of the previous section that all $D$ terms disappear in the
double scaling limit. There it was also found that all $J$-dependent $A$
terms would vanish. This also appears from the formulas above. No
$J$ terms ever appear if we do not start out with any and we do not.

 The starting point of the iteration procedure is of course as before
$W_0(p,p)$ but now we should take only the part of it that contributes
in the double scaling limit. From equation~\rf{twoloop} we
find\footnote{Here and in the following we use the notation that the
superscript $(S)$ refers to the case of the symmetric potential and the
superscript $(NS)$ to the case of the non-symmetric potential (where the
critical behaviour is associated with the endpoint $x$).}
\beq
W_0^{(NS)}(p,p)=\frac{1}{16}\frac{1}{(p-x)^2} \hspace{0.7cm} (d.s.l.).
\eeq
There is an interesting feature of the $A$ coefficients.
For all genera it holds
that $A_g^{(1)}=A_g^{(2)}=0$. This is  because the smallest negative
power of $(p-x)$ appearing in any basis vector is 3/2. By multiplication
of two basis vectors or by application of the loop insertion operator
this power will be lowered by at least 3/2. Hence terms of the type
$(p-x)^{-1}$ and $(p-x)^{-2}$ never turn up on the right hand side of
the loop equation.

We have calculated the $A$ coefficients as they look
in the double scaling limit for $g=2$, 3 and~4 following the iterative
procedure outlined above. The results for $g=2$ and $g=3$ read
\begin{eqnarray}
A_2^{(1)}& = & 0 \;,
      \nonumber \\
A_2^{(2)}&= & 0 \;,
     \nonumber \\
A_2^{(3)} &= &
{{49\,{{M_{2}}^2}}\over {256\,d_c\,{{M_{1}}^4}}} -
   {{5\,M_{3}}\over {32\,d_c\,{{M_{1}}^3}}} \;,
     \nonumber \\
A_2^{(4)} &=&
{{-49\,M_{2}}\over {128\,d_c\,{{M_{1}}^3}}} \;,
\nonumber \\
A_2^{(5)} &=&
{{105}\over {256\,d_c\,{{M_{1}}^2}}} \;,
\nonumber \\
A_3^{(1)}& = & 0 \;,
      \nonumber \\
A_3^{(2)}&= & 0 \;,
     \nonumber \\
A_3^{(3)} &= &
  {{-5355\,{{M_{2}}^5}}\over {512\,{d_c^2}\,{{M_{1}}^9}}} +
   {{7995\,{{M_{2}}^3}\,M_{3}}\over {256\,{d_c^2}\,{{M_{1}}^8}}} -
   {{32845\,M_{2}\,{{M_{3}}^2} + 35588\,{{M_{2}}^2}\,M_{4}}\over
     {2048\,{d_c^2}\,{{M_{1}}^7}}}
\nonumber\\*
 &&+ {{21\,\left( 680\,M_{3}\,M_{4} + 773\,M_{2}\,M_{5} \right) }\over
     {2048\,{d_c^2}\,{{M_{1}}^6}}} -
   {{1155\,M_{6}}\over {512\,{d_c^2}\,{{M_{1}}^5}}} \;,
\nonumber \\
A_3^{(4)} &= &
{{5355\,{{M_{2}}^4}}\over {256\,{d_c^2}\,{{M_{1}}^8}}} -
   {{21837\,{{M_{2}}^2}\,M_{3}}\over {512\,{d_c^2}\,{{M_{1}}^7}}}
\nonumber \\*
 &&+   {{17545\,{{M_{3}}^2} + 37373\,M_{2}\,M_{4}}\over
     {2048\,{d_c^2}\,{{M_{1}}^6}}} -
   {{10185\,M_{5}}\over {2048\,{d_c^2}\,{{M_{1}}^5}}}
 \;, \nonumber \\
A_3^{(5)} &= &
{{-7371\,{{M_{2}}^3}}\over {256\,{d_c^2}\,{{M_{1}}^7}}} +
   {{69373\,M_{2}\,M_{3}}\over {2048\,{d_c^2}\,{{M_{1}}^6}}} -
   {{17465\,M_{4}}\over {2048\,{d_c^2}\,{{M_{1}}^5}}} \;,
\nonumber \\
A_3^{(6)} &= &
{{64295\,{{M_{2}}^2}}\over {2048\,{d_c^2}\,{{M_{1}}^6}}} -
   {{30305\,M_{3}}\over {2048\,{d_c^2}\,{{M_{1}}^5}}} \;,
\nonumber \\
A_3^{(7)} &= &
{{-13013\,M_{2}}\over {512\,{d_c^2}\,{{M_{1}}^5}}} \;,
\nonumber \\
A_3^{(8)} &= &
{{25025}\over {2048\,{d_c^2}\,{{M_{1}}^4}}}\;.
\end{eqnarray}
We remind the reader that the terms listed above are only potentially
relevant. Whether or not they are actually relevant depends on which
multi-critical model one wants to consider. For a $m$'th  multi-critical
model all terms involving $M_k$, $k>m$ vanish in the double scaling limit.
This is true whether one uses a minimal potential or not.
We also remind the reader that we assumed that we had a non-symmetric
potential and that the critical behaviour was associated with the
endpoint $x$. In the case where the critical behaviour is associated
with the endpoint $y$ all the formulas in this section
still hold provided $d_c$ is replaced with $-d_c$, $x$
with $y$ and $M$ with $J$.

Let us turn now to the case of the symmetric potential. In view of the
scaling relations~\rf{scalps} and~\rf{scalzs} it might seem unnatural
to work with terms like $(p-x)$ and $(p-y)$ and we will see below that
there  actually exists a way of avoiding this. However, for the moment
we will analyze the scaling behaviour
in the $x,y$ formalism by (formally) scaling $p$ to $-\sqrt{z_c}$
whenever it occurs in a term like $(p-y)$ and to $+\sqrt{z_c}$ whenever
it occurs in a term like $(p-x)$. Now let us assume as in the
non-symmetric case that we know the scaling relevant versions of
$A_G^{(n)}$ and $D_G^{(n)}$ for $G=1,\ldots,g$ and
that we want to calculate the double scaling limit of the right hand
side of the loop equation for $W_{g+1}(p)$. Of course we still have that
all terms in the basis vectors show the same scaling behaviour so we
still have to keep all of them. However it is easy to convince oneself that
in the double scaling limit all occurrences of $(p-y)$ in $\chi^{(n)}(p)$
can be replaced by $d_c$ and all occurrences of $(p-x)$ in $\Psi^{(n)}$
can be replaced by $-d_c$. Furthermore, it appears that when we calculate
the sum of products on the right hand side of the loop equation all
products that mix $\chi$'s and $\Psi$'s will be subdominant in the
double scaling limit. These mixed products should hence be discarded.
Finally to ensure the survival of only double scaling relevant terms in
$W_g(p,p)$, the loop insertion operator should be written as
\beq
\frac{d}{dV(p)}=\frac{d}{dV(p)}_{x}+\frac{d}{dV(p)}_{y}
\eeq
where $d/dV(p)_y$ can be obtained from $d/dV(p)_x$ given in
equation~\rf{ddVx1} -- \rf{ddVx4} by replacing $x$ by $y$, $M$ by $J$
and $d_c$ by $-d_c$.

The starting point for the iterative procedure is of
course the double scaled version of $W_0(p,p)$ which here reads
\beq
W_0^{(S)}(p,p)=\frac{1}{16}\frac{1}{(p-x)^2}
+\frac{1}{16}\frac{1}{(p-y)^2} \hspace{0.7cm} (d.s.l.).
\eeq
We see that the loop equation actually decouples into two independent
equations. Each of these is a double scaled version of the loop equation
for the non symmetric potential. One corresponds to the case where the
critical behaviour is associated with the endpoint $x$, the other to
the case where the critical behaviour is associated with the endpoint
$y$. The $A_g^{(n)}$'s are hence equal to those obtained in the
non-symmetric case where the critical behaviour is associated with the
endpoint $x$ and $D_g^{(n)}=(-1)^n A_g^{(n)}$.
This means that the right hand side of the loop equation of the loop
equation can be written as
\beq
\hbox{the r.h.s.}=\sum_{n=1}^{3g-1} d_c^n A_g^{(n)} \frac{1}{(p^2-z)^n}
\label{rhsform}
\eeq
which seems to be the natural way of expressing it  bearing in mind the
scaling relations~\rf{scalps} and~\rf{scalzs}.

As a consequence, it
becomes also more natural to express the generating functional as
\beq
W^{(S)}_g(p) = \sum_{n=1}^{3g-1}\tilde{A}^{(n)}_g \tilde{\chi}^{(n)}(p)
\eeq
where
$\tilde{A}_g^{(n)}=A_g^{(n)} d^n_c$ and where we now have only one set
of basis vectors $\tilde{\chi}^{(n)}(p)$ which fulfill
\beq
\left\{\hat{K}-2W_0(p)\right\} \tilde{\chi}^{(n)}(p)=\frac{1}{(p^2-z)^n}.
\eeq
It is easy to show that $\tilde{\chi}^{(n)}(p)$ should be chosen as
\beq
\tilde{\chi}^{(n)}(p)=\frac{1}{M_1}
\left\{\Phi_z^{(n)}(p)-\sum_{k=1}^{n-1}\tilde{\chi}^{(k)}(p)M_{n-k+1}
d_c^{k-n}\right\}
\eeq
where
\beq
\Phi_z^{(n)}(p)=\frac{1}{(p^2-z)^{n+1/2}}
.\eeq

Furthermore, one can easily convince oneself that the right hand side of
the loop equation can be obtained
directly in the form~\rf{rhsform} if one  replaces $(p-x)^{-1}$ by
$d_c(p^2-z)^{-1}$ in the formulas~\rf{ddVx1} -- \rf{ddVx4}, \ie if one carries
out the iteration process starting from
\beq
W^{S}_0(p,p)=\frac{d_c^2}{16(p^2-z)^2}\label{Hdsl1}
\hspace{0.7cm} (d.s.l.)
\eeq
and using the following expression for the loop insertion operator.
\beq
\frac{d}{dV_s(p)}=\sum_n \frac{dM_n}{dV(p)}\frac{\partial}{\partial M_n}
+\frac{dz}{dV(p)}\frac{\partial}{\partial z} \label{Hdsl2}
\hspace{0.7cm}(d.s.l.)
\eeq
where
\bea
\frac{dM_n}{dV(p)}&=&-(n+1/2)\left\{\Phi_z^{(n+1)}(p)d_c^n
-\frac{M_{n+1}}{M_1}\Phi_z^{(1)}(p)\right\}d_c
\hspace{0.7cm} (d.s.l.)\,,
\label{Hdsl3}\\
\frac{dz}{dV(p)}&=&\frac{d_c}{M_1}\Phi_z^{(1)}(p)
\hspace{0.7cm}(d.s.l.).
\label{Hdsl4}
\eea
This observation allows us t show explicitly that the hermitian and the
complex matrix model are equivalent in the double scaling limit.

\subsection{Determining $F_g$ in the double scaling limit \label{F_g}}

The starting point of the calculation is the double scaled
version of $W_g(p)$ obtained as described in the previous section
--- and the strategy is again to rewrite the basis vectors in a form which
allows one to identify $W_g(p)$ as a total derivative. However this time
the rewriting in the case of $\chi^{(n)}$
is made with the aid of~\rf{ddVx2} instead of~\rf{dMdV} and reads
\beq
\chi^{(n)}(p)=\frac{1}{M_1}\left\{
\frac{-2}{2n-1}\frac{dM_{n-1}}{dV(p)}
-\sum_{k=2}^{n-1}\chi^{(k)}(p)M_{n-k+1}\right\}
\hspace{0.7cm} n\geq 2\;.
\label{chinew}
\eeq
We do not need any expression for $\chi^{(1)}(p)$ since the sum
in~\rf{chinew} starts at $k=2$ and since we know that in the
double scaling limit $A_g^{(1)}=0$, neither we do need any expression
for $\Phi_x(p)$.
The relevant rewriting of $\Psi^{(n)}(p)$ appears from~\rf{chinew} when
$M$ is replaced with $J$.

 For the non-symmetric potential where the
critical behaviour is associated with the endpoint $x$, we immediately
find for genus one
\beq
F^{(NS)}_1=-\frac{1}{24}\ln M_1 \hspace{0.7cm} (d.s.l.).
\eeq
This expression can of course alternatively be obtained by taking
\dsl\ of \eq{hermF1}.

Before presenting the explicit results for $F_g^{(NS)}$ in genus $2,3$ and
$4$, let us describe the general structure of $F_g^{(NS)}$ as it
appears from our iterative solution of the loop equation.
We have
\beq
F_g^{(NS)}=\sum_{\a_j>1}\langle \a_1 \ldots \a_s |\a,
\g \rangle_g^{herm}
{M_{\a_1}\ldots M_{\a_s}\over M_1^{\a}d^{\g}}
\hspace{0.5cm}(d.s.l.)\hspace{0.5cm} g \geq1\
\label{fgendsl}
\eeq
where the rational numbers denoted by the brackets are identical to those of
\eq{fgen} having the same indices.
The $\a_j$'s obey \eq{homas}, $\g$ is given by \eq{g-1} and
\be
\a=2g-2+s
\label{ag}
\ee
as it follows from \eq{NMas}.

Let us consider the $g=2$ case as an example of what structures are encoded
by \eq{fgendsl}. For $g=2$ we have $\a_j\in[2,3g-2]=[2,4]$ and the
relation~\rf{homas} gives
\be
\sum_{j=1}^s(\a_j-1)=3g-3=3 \hspace{1.5cm}g=2\;,
\label{homasg=2}
\ee
so that $s\in[1,3g-3]=[1,3]$. The restriction~\rf{homasg=2} admits
$s=3$ and $\a_1=\a_2=\a_3=2$, $s=2$ and ${\a_1,\a_2}={2,3}$
and $s=1$ and $\a_1=4$. The proper powers
of $M_1$ in the denominator for such terms are given by $\a=2g-2+s=2+s$.
The outcome of the iteration process is
\be
F_2^{(NS)}=
{{-21\,{{M_2}^3}}\over {160\,d_c\,{{M_1}^5}}} +
   {{29\,M_2\,M_3}\over {128\,d_c\,{{M_1}^4}}} -
   {{35\,M_4}\over {384\,d_c\,{{M_1}^3}}}\;.
\label{F2}
\ee
We note that all allowed values of $s$ and $\a_i$'s actually appear. By
taking a closer look at the loop insertion operator and the basis
vectors it is easy to convince oneself that this will be the case for
all genera. The expression~\rf{F2} could of course also have been obtained
by taking the double scaling limit of~\rf{f2complete}
 following the recipe given
in~\rf{NMas} and~\rf{homas}. We see that the expression for $F_2$ simplifies
drastically in the double scaling limit.

The results for $g=3$ (where $s\leq6$) and $g=4$ (where $s\leq9$) read
\bea
F_3^{(NS)}&=& {{2205\,{{M_2}^6}}\over {256\,{d_c^2}\,{{M_1}^{10}}}} -
   {{8685\,{{M_2}^4}\,M_3}\over {256\,{d_c^2}\,{{M_1}^9}}} +
   {{15375\,{{M_2}^2}\,{{M_3}^2}}\over {512\,{d_c^2}\,{{M_1}^8}}}
+{{5565\,{{M_2}^3}M_4}\over {256\,{d_c^2}\,{{M_1}^8}}}
\nonumber\\* &&
    - {{5605\,M_2\,M_3 \,M_4}\over {256\,{d_c^2}\,{{M_1}^7}}}
   -{{72875\,{{M_3}^3}}\over {21504\,{d_c^2}\,{{M_1}^7}}}
-{{ 3213\,{{M_2}^2}\,M_5}\over{256\,{d_c^2}\,{{M_1}^7}}}
+{{ 2515\,M_3\,M_5}\over{512\,{d_c^2}\,{{M_1}^6}}}
\nonumber\\*
&& +
 {{21245\,{{M_4}^2}}\over {9216\,{d_c^2}\,{{M_1}^6}}}
   + {{5929\,M_2\,M_6}\over {1024\,{d_c^2}\,{{M_1}^6}}} -
    {{5005\,M_7}\over {3072\,{d_c^2}\,{{M_1}^5}}}
\eea
and
\bea
F_4^{(NS)}&=&
-{{21023793\,{{M_2}^9}}\over {10240\,{d_c^3}\,{{M_1}^{15}}}}
+    {{12829887\,{{M_2}^7}\,M_3}\over {1024\,{d_c^3}\,{{M_1}^{14}}}}
-   {{98342775\,{{M_2}^5}\,{{M_3}^2}}\over
     {4096\,{d_c^3}\,{{M_1}^{13}}}}\nonumber
\\* &&
 - {{4456305\,{{M_2}^6}\,M_4}\over {512\,d_c^3\,{{M_1}^{13}}}}
  + {{16200375\,{{M_2}^3}\,{{M_3}^3}}\over
      {1024\,{d_c^3}\,{{M_1}^{12}}}}
+ {{26413065\,{{M_2}^4}\,M_3\,M_4}\over{1024\,d_c^3\,{{M_1}^{12}}}}
\nonumber \\* &&
+  {{12093543\,{{M_2}^5}\,M_5}\over {2048\,d_c^3\,{{M_1}^{12}}}}
-     {{83895625\,M_2\,{{M_3}^4}}\over {32768\,{d_c^3}\,{{M_1}^{11}}}}
- {{68294625\,{{M_2}^2}\,{{M_3}^2}\,M_4}\over
  {4096\,d_c^3\,{{M_1}^{11}}}} \nonumber \\ &&
- {{12367845\,{{M_2}^3}\,{{M_4}^2}}\over
     {2048\,d_c^3\,{{M_1}^{11}}}}
-   {{13024935\,{{M_2}^3}\,M_3\,M_5}\over
     {1024\,d_c^3\,{{M_1}^{11}}}}
-    {{15411627\,{{M_2}^4}\,M_6}\over {4096\,d_c^3\,{{M_1}^{11}}}}
\nonumber\\ &&
+  {{32418925\,{{M_3}^3}\,M_4}\over {24576\,d_c^3\,{{M_1}^{10}}}}
+   {{17562825\,M_2\,M_3\,{{M_4}^2}}\over
     {4096\,d_c^3\,{{M_1}^{10}}}}
+    {{578655\,M_2\,{{M_3}^2}\,M_5}\over
     {128\,d_c^3\,{{M_1}^{10}}}} \nonumber \\*
&&
+    {{10050831\,{{M_2}^2}\,M_4\,M_5}\over
     {2048\,d_c^3\,{{M_1}^{10}}}}
+    {{5472621\,{{M_2}^2}\,M_3\,M_6}\over
     {1024\,d_c^3\,{{M_1}^{10}}}}
+    {{44207163\,{{M_2}^3}\,M_7}\over
     {20480\,d_c^3\,{{M_1}^{10}}}}
\nonumber \\*
&&
-    {{1511055\,M_2\,{{M_5}^2}}\over {2048\,{d_c^3}\,{{M_1}^9}}}
-  {{7503125\,{{M_4}^3}}\over {36864\,{d_c^3}\,{{M_1}^9}}}
-    {{2642325\,M_3\,M_4\,M_5}\over {2048\,d_c^3\,{{M_1}^9}}}
\nonumber \\*
&&
-    {{11532675\,{{M_3}^2}\,M_6}\over {16384\,d_c^3\,{{M_1}^9}}}
-    {{6242775\,M_2\,M_4\,M_6}\over {4096\,d_c^3\,{{M_1}^9}}}
-    {{6968247\,M_2\,M_3\,M_7}\over {4096\,d_c^3\,{{M_1}^9}}}
\nonumber \\* &&
-    {{4297293\,{{M_2}^2}\,M_8}\over {4096\,d_c^3\,{{M_1}^9}}}
+   {{12677665\,M_2\,M_9}\over {32768\,{d_c^3}\,{{M_1}^8}}}
+    {{8437275\,M_5\,M_6}\over {32768\,d_c^3\,{{M_1}^8}}}
\nonumber \\*
&&
+    {{8913905\,M_4\,M_7}\over {32768\,d_c^3\,{{M_1}^8}}}
+    {{10156575\,M_3\,M_8}\over {32768\,d_c^3\,{{M_1}^8}}}
-    {{8083075\,M_{10}}\over {98304\,{d_c^3}\,{{M_1}^7}}}\;.
\eea
Needless to say that the results for the case where the critical
behaviour is associated with the endpoint $y$ can be obtained from these
by performing the replacements $M\rightarrow J$, $d_c\rightarrow -d_c$.

It is obvious from the discussion in Section~\ref{dsl} and
formula~\rf{chinew} that independently of the details of the
prescription for taking the double scaling limit,  $F_g$
depends in this limit on at most $3g-2$ moments. Again we stress that
the terms listed above are only potentially relevant. For a $m$'th
multi-critical model all terms involving $M_k$, $k>m$ vanish too. This
is true whether one uses a minimal potential or not.
It is interesting to note that for the symmetric potential we get a sum
of two identical terms and hence
\beq
F_g^{(S)}=2F_g^{(NS)} \hspace{0.7cm} (d.s.l.)
\label{twice}
\eeq
which is a well known property of \dsl\ for \hm~\cite{BP90}.
It can be traced back to the fact that the loop equation for a symmetric
potential decouples into two independent equations.
Equation~\rf{twice} is the reason why
the partition function of 2D quantum gravity is defined as the
square root of \dsl\ of the partition function~\rf{partition}
in the case of a symmetric potential. There is no such square root in the
non-symmetric case and the proper continuum partition function can be
obtained just as \dsl\ of~\rf{partition}.

Of course we have determined up till now only the coefficients, $F_g$, of the
genus expansion of the free energy (Cf. Eq~\rf{freeen}). For a $m$'th
multi-critical model the relevant expansion parameter in the double
scaling limit is $Na^{m+1/2}$. If we define moments $\mu_k$ by
(Cf. Eq~\rf{scalmom})
\beq
M_k=a^{m-k}\mu_k,\hspace{0.5cm}k\in[1,m]
\eeq
we get by replacing $M_k$ with $\mu_k$ for $k\in [1,m]$
and setting $M_k$ equal to zero for $k>m$ in the formulas above exactly
the coefficient of the expansion in the double scaling parameter. A
similar statement holds for the results of the previous section.

\subsection{A remark on the complex matrix model \label{complex} }

The complex matrix model is defined by the partition function
\beq
Z^{C}=e^{N^2F^C}=
\int_{N\times N} d\phi^{\dg} d\phi\exp(-NV_C(\phi^{\dg} \phi))
\label{ZC}
\eeq
where the integration is over complex $N\times N$ matrices and
\beq
V_{C}(\phi^{\dg} \phi)=\sum_{j=1}^{\infty}\frac{g_j}{j}\tr(\phi^{\dg}\phi)^j\,.
\eeq
Its generating functional is defined by
\beq
W^C(p)=\frac{1}{N}\sum_{k=0}^{\infty}\langle \tr(\phi^{\dg}\phi)^{k}\rangle
/p^{2k+1}
\eeq
and the $s$-loop correlator $(s\geq 2)$ by
\beq
W^C(p_1,\ldots,p_s)=N^{s-2}\sum_{k_1,\ldots,k_s=1}^{\infty}
\langle \tr(\phi^{\dg}\phi)^{k_1} \ldots
\tr(\phi^{\dg}\phi)^{k_s})\rangle_{conn.}
/p_1^{2k_1+1}\ldots p_s^{2k_s+1}
\eeq
so that
\be
W^C(p_1,\ldots,p_s)=N^{s-2} \LA  \tr \frac{p_1}{p_1^2-\phi^\dg\phi}
\ldots\tr \frac{p_s}{p_s^2-\phi^\dg\phi} \RA_{conn.}\,.
\ee

In Ref.~\cite{complex} an iterative procedure which enables one to calculate
$W_g^C(p_1,\ldots,p_s)$ for any genus $g$ and any $s$ starting from
$W_0^C(p)$ was developed. This procedure is much the same as for the
hermitian matrix model. It is based on the loop
equation, the loop insertion operator and a suitable change of variables.
In Ref.~\cite{complex} the studies were carried out only away from the double
scaling limit. In  the Appendix we show how the iterative scheme for the
complex matrix model can be modified to give directly the result in the
double scaling limit. The modified scheme provides us with an exact
proof that the hermitian and the complex matrix model are equivalent in
the double scaling limit%
\footnote{The fact that the hermitian and complex matrix models
belong to the same universality class and that the
correlators therefore coincide in \dsl\ does not contradict
Ref.~\cite{MMMM} where it was shown  that the discrete Virasoro
operators for the complex matrix model do not have continuum limit.}.
We find that
\bea
W_g^{C}(p_1,\ldots,p_s) &=& \frac{1}{4^{g+s-1}}W_g^{(S)}(p_1,\ldots,p_s)
\hspace{0.7cm}(d.s.l.) \,,\\
F_g^C&=&\frac{1}{4^{g-1}}F_g^{(S)}\hspace{0.7cm}(d.s.l.)
\eea
independently of the type of critical behaviour considered.
In particular we see that the partition function of 2D quantum gravity
can also be obtained as the square root of the double scaling limit of
the partition function~\rf{ZC} provided the integration is over
$N/2\times N/2$ matrices in stead of over $N\times N$ matrices.
This is of course in accordance with the fact that the number of
independent components of a $N/2\times N/2$ complex matrix is the same
as that of a $N\times N$ hermitian matrix.


\newsection{Relation to moduli space}

The equivalence between \dsl\ of \hm\ and the Kontsevich model has been
proven in various ways in Refs.~\cite{W3,MMM}.
We present in this section another proof directly in terms of the moments
introduced above. For this purpose we describe a limiting
procedure which allows us to obtain the
Kontsevich model directly from \hm.
We discuss also the connection between \hm\
and a certain discretization of moduli space.

\subsection{The limiting procedure \label{l.p.}}

As we have seen in Section~\ref{F_g}, \dsl\ of the free energy
is given by the expansion~\rf{fgendsl} which encodes in a remarkable
way all multi-critical behaviour of \hm.
Since most of the terms originally present in the expression for $F_g$
vanish in the double scaling limit the
 question arises whether it is possible to formulate a matrix model
which gives solely the coefficients which are relevant for \dsl\.
The answer to this question is --- yes
--- and the proper model is known as the Kontsevich model.


The Kontsevich model is
defined by the integral~\cite{Kon91}
\beq
Z_K[M,n]={\int_{n\times n}dX \exp\left\{ n\str \left(-\frac 12 MX^2+
\frac{X^3}{6}
\right)\right\} \over
\int_{n\times n}dX \exp\left\{ n\str \left(-\frac 12 MX^2\right)\right\} }\;.
\label{Kont}
\eeq
there $M$=diag$(m_1,\ldots,m_n)$ (at first)
is a positive definite matrix and the integration
is over hermitian $n\times n$ matrices.
The genus expansion of the free energy is given by
\be
\ln Z_K[M,n]=\sum_{g=0}^\infty n^{2-2g} F_g^{Kont}\,.
\label{freeenkont}
\ee

The goal of this section is to show that the Kontsevich model can be
obtained from \hm\ by a certain {\it limiting procedure\/} which looks
conceptually different from \dsl. We shall see however in
Section~\ref{lkontsevich} that this limiting procedure leads to a
result which is identical to the one obtained in \dsl.

In order to formulate the limiting procedure
let us recall some identities.
Consider the hermitian matrix model as defined in \eq{partition}
with $N = -\baA n$.%
\footnote{We should point out
a subtlety related to the representations \rf{kp} and \rf{kp1}.
We have used a constant $\baA$ in these equations which has the opposite
sign of that conventionally used. The ``analytic'' continuation of
$\baA$ is needed in order to make contact with the Kontsevich model.
We will not to try to ``justify'' it, since the idea
is to provide some heuristic explanation of the observed identity
of the expansion coefficients of the hermitian matrix model and the
Kontsevich model.}
We have the following identity \cite{cm}%
\footnote{See also Ref.~\cite{KMMM} where the proof was extended to
finite $n$.}
\be
Z[g,N(\baA)]= \e^{-n\str \Lambda^2/2}
Z_P[\Lambda,n], \hspace{1.7cm} N(\baA)=-\baA n
\label{oo}
\ee
where the partition function $Z_P[\LL,n]$ is
associated with the Penner model~\cite{Pen},
 in the external field $\LL=$diag$(\lL_1,\dots,\lL_n)$:
\beq
Z_{P}[\LL,n]=\int_{n\times n} dX \exp\left[n\str \left(\LL X-\frac 12 X^2-
\baA \log X\right)\right]. \label{kp}
\eeq
The identification is valid provided the set of coupling constants
$\{ g \}=\{g_0,g_1,\ldots\} $ and the matrix $\LL$
are related by the Miwa transformation
\beq
g_k= \frac 1n\str\LL^{-k}-\delta_{k2}\hspace{1.5cm}\ k\geq 1,\quad
g_0=\frac 1n \str\log \LL^{-1} .\label{tt}
\eeq

In order to show how the partition function~\rf{Kont} can be obtained
from~\rf{kp},
we note that $Z_P [\LL,n]$ can be
reformulated as the so-called Kontsevich--Penner model~\cite{cm1}:
\beq
Z_{KP}[\h,n]={\int_{n\times n}dX \exp\left\{-\baA n\str\left[ \frac 12\h X\h X
+(\log(1-X)+X)\right]\right\} \over
\int_{n\times n}dX \exp\left\{-\baA n\str\left[\frac 12\h X\h X-\frac 12
X^2\right]\right\} }\label{kp1}
\eeq
provided
\beq
\LL = \sqrt{\baA}(\h +\h^{-1}).
\label{add6}
\eeq
We note in addition
that the model \rf{kp1} possesses a remarkable connection to
a discretization of moduli space which will be discussed in
Section~\ref{disk}.
Here we will be interested in the connection to the usual continuum
moduli space, \ie to the Kontsevich model which precisely has the
interpretation as  the generating functional for intersection indices
on moduli space (Cf. to Section~\ref{i.i.}).
In order to highlight this connection we choose
a specific parametrization of $\h$ and $\baA$:
\beq
\h = \e^{\eps m},~~~~~~\baA = \frac{1}{2\eps^3}
\label{add7}
\eeq
where $m$ is to be identified with the matrix, $M$ which appears
in the definition \rf{Kont} of the Kontsevich model, and where $\eps$
is presently just an expansion parameter which we are going to take
to zero, but which in Section~\ref{disk}
will be given the interpretation of the
step of discretization of moduli space.

It is now trivial to check that  we in the limit $\eps \to 0$
reproduce \rf{Kont} from \rf{kp1} after rescaling $X\ra \eps X$.
In this way we have managed to
move continuously from $Z[g, N(\baA ) ]$ to $Z_K [m,n]$.
Note that there is no need to take the limit $n \to \infty$ during
these steps, but the size $N(\baA)= -\baA n$ of the original hermitian
matrix goes to infinity as $\eps \to 0$.

It is instructive to illustrate the proof just given
by taking the $\eps\ra0$ limit of the explicit formulas governing the
hermitian matrix model.
The basic equations \rf{xandy} which in the original matrix model determined
the endpoints $x$ and $y$ in terms of the coupling constants $g_i$
have an equivalent formulation in terms of the eigenvalues $\l_j$ of
$\LL$ when we rewrite the matrix model in terms of the Penner model
as in \rf{oo}. They are obtained by inserting in \rf{xandy} the
following relation between the eigenvalues $\l_j$ and the matrix model
potential, $V$:
\beq
V^\prime(\omega)=\frac 1n \sum_{j=1}^n\frac{1}{\lL_j-\omega}-\omega.
\label{rho}
\eeq
This relation follows directly from \rf{tt}. By the use of \rf{rho}
the equation \rf{xandy} reads
\bea
\frac 1n \sum_{i=1}^n {1\over \sqrt{(\l_i-x)(\l_i-y)}}- \frac{x+y}{2} &=&
0\,,
\label{e1}\\
\frac 1n \sum_{i=1}^n {\l_i-\frac {x+y}{2}\over \sqrt{(\l_i-x)(\l_i-y)}}
- \frac{(x-y)^2}{8} &=& -2\baA.
\label{e2}
\eea
Notice that  the normalization factor ``2'' in \rf{xandy} has been changed
to $-2\baA$, since the size of the matrix is $N = -\baA n$.

Let us now apply the limiting procedure dictated by \rf{add7} to
Eqs.~\rf{e1} and \rf{e2}:
\beq
\Lambda=\sqrt{\baA}(\e^{\eps m}+\e ^{-\eps m})=
\frac {\sqrt{2}}{ \eps^{3/2}}+\sqrt{\frac{\eps}{2}} m^2
+O(\eps^{5/2}). \label{eps}
\eeq
In order to  solve equations \rf{e1}, \rf{e2} in terms of $x$ and $y$
we choose
\be
y=-\frac{\sqrt{2}}{\eps^{3/2}}\,,
\hspace{1cm} x=\frac{\sqrt{2}}{\eps^{3/2}}+
\sqrt{2}\eps^{1/2}  u_0 +\ldots\;.
\label{epsxy}
\ee
{}From \eq{e1} we have the equation
\beq
\frac 1n \sum_{j=1}^n {1\over\sqrt{m_j^2-2u_0}}=u_0 \label{add5}
\eeq
which appears multiplied by $\sqrt{\eps}$.
After the substitution into \eq{e2}, we get that the leading term of order
$1/\eps^3$ coming from $2\baA$ and $(x-y)^2/8$ cancels. It is crucial
for this cancellation that $\baA>0$.
The next order
terms which are proportional to $1/\eps$ combine again into \eq{add5},
which we recognize as nothing but the stationary condition~\cite{MS}
 appearing in the Kontsevich model.

Let us now demonstrate how the $1$-loop average of \hm\ is related to the
one of  the Kontsevich model. We start from genus zero inserting
\eq{rho} into \rf{zero} which gives~\cite{cm}
\be
W_0(p)=\frac 12 \left[\sqrt{(p-x)(p-y)}-p
+\frac 1n \sum_j \frac{1}{p-\l_j}\left(
\frac{\sqrt{(p-x)(p-y)}}{\sqrt{(\l_j-x)(\l_j-y)}}-1
\right)\right]\,.
\label{5.15}
\ee
Eq.~\rf{e1} ensures the vanishing of the ${\cal O}(1)$ term of the $1/p$
expansion of the right hand side, while \eq{e2} implies that the
${\cal O}(p^{-1})$ term equals  $-\baA /p$. Contrary the
representation~\rf{zeroint} can be obtained from~\rf{5.15} by expanding in
$1/\l_j$ and using Eqs.~\rf{tt} and \rf{rho}.

To obtain an analogue of \eq{eps} for $p$, let us
point out that the relation~\rf{tt} implies
\be
\frac{\d}{\d\l_i}= -\frac {1}{\l_i}\frac{\d}{\d g_0}
-\sum_{k=1}^\infty \frac{k}{\l_i^{k+1}}
\frac{\d}{\d g_k} \hspace{1.5cm} n\ra\infty
\ee
which requires $n\ra\infty$ in order for $g_k$'s to be
independent variables. Comparing with \eq{2.8} one gets
\be
\frac{d}{dV(p)}=\frac{\d}{\d\l_i}\Big|_{\l_i=p}
\hspace{1.5cm} n\ra\infty\,.
\label{5.17}
\ee
Therefore we substitute
\beq
p=
\frac {\sqrt{2}}{ \eps^{3/2}}+\sqrt{\frac{\eps}{2}} \pi
+O(\eps^{5/2}) \label{ppi}
\eeq
where $\pi$ is to be understood as the momentum variable for
the Kontsevich model.

Substituting \rf{eps}, \rf{epsxy} and \rf{ppi} into \eq{5.15}, one gets
\be
W_0(p)+\frac p2 +\frac{1}{2n}\sum_j\frac{1}{p-\lambda_j}
\ra
\sqrt{\frac 2\eps}\left[ W_0^{kont}(\pi)+\frac{\sqrt{\pi}}{2}
+\frac{1}{2n}\sum_k \frac{m_k}{\sqrt{\pi}}\,\frac{1}{\pi-m_k^2}\right]
\label{renorm}
\ee
where
\bea
W_0^{kont}(\pi)&=&
\frac{\d}{\d m_i^2} \Big|_{m_i^2=\pi}\ln{ Z_0^{Kont}[M,n]} \nonumber\\
&=&
\frac 12 \left[\sqrt{\pi-2u_0}-\sqrt{\pi}
+\frac 1n \sum_j \frac{1}{\pi-m_j^2}\left(
\frac{\sqrt{\pi-2u_0}}{\sqrt{m^2_j-2u_0}}-\frac{m_j}{\sqrt{\pi}}
\right)
\right]
\label{5.20}
\eea
We see that the subtraction needed to make contact with the Kontsevich
model is exactly the same as the usual double scaling subtraction.
For the two-loop correlator we have
\beq
W_0(p,q)+\frac{1}{2}\frac{1}{(p-q)^2}\ra
\left(\frac{2}{\epsilon}\right)
\left[W_0^{Kont}(\pi_1,\pi_2)+\frac{\frac{1}{2}(\pi_1+\pi_2)}
{2(\pi_1-\pi_2)^2\sqrt{\pi_1\pi_2}}\right]
\label{pq}
\eeq
We recognize the term $1/2(p-q)^2$ as being the non universal part of
the two-loop correlator. Hence the renormalization needed in this case
is also the usual one of the double scaling limit. For $p=q$ we
find
\beq
W_0(p,p)\ra
\left(\frac{2}{\epsilon}\right)
\left[W_0^{Kont}(\pi,\pi)+\frac{1}{16\pi^2}\right]
\label{2loop}
\eeq
The appearance of the term $1/16\pi^2$ when one wants to make contact
with continuum physics is well known from the study of the double
scaling limit of the Virasoro constraints for the hermitian matrix
model~\cite{MMMM}.
It is evident from equation~\rf{pq} that for the higher
loop correlators it holds that
\beq
W_0(p_1,\ldots,p_s) \ra \left(\frac{2}{\epsilon}\right)^{s/2}
W_0^{Kont}(\pi_1,\ldots,\pi_s),\hspace{0.5cm} s>2
\label{higdsl}
\eeq
This multiplicative renormalization is also in compliance with the
renormalization of the double scaling limit. The origin of the factors
$(2/\epsilon)^{1/2}$ is easy to understand bearing in mind the
relations~\rf{eps} and~\rf{5.17}. In the next section we show that the
limiting procedure described above actually leaves us with the same
terms as the double scaling limit does. It is worth noting that this
means that the Kontsevich correlators encode information about continuum
correlators at all multi-critical points. This
information is even easily accessible.
To obtain the continuum correlators
one rewrites first \eq{add5} as the genus zero string equation
\be
\sum_{k=0}^\infty \frac{(2k-1)!!}{k!}t_k u_0^k =0
\label{se0}
\ee
where
\be
t_k=\frac 1n \str M^{-2k-1} -\delta_{k1} \hspace{1.7cm} k\geq 0
\ee
are continuum coupling constants which all vanish at the $m^{th}$
multi-critical point except for  $t_m$ and $t_0$ with the latter
playing the role of the cosmological constant. The continuum
correlators can hence be found by expanding the Kontsevich correlators
in powers of $1/m_j$. Equation~\rf{add5} ensures the vanishing of the
$1/\sqrt{\pi}$ term in the the expansion of the right hand side
of~\rf{5.20}.

We are now ready to present one more proof of the fact that by carrying our
limiting procedure for \hm\ we recover the Kontsevich model. For this aim
let us consider the Schwinger--Dyson equation of the model~\rf{kp}
which reads~\cite{cm1}
\beq
 {\left. \frac{1}{n^2}\frac{\d \tilde W(\l_i)}{\d
\l_j}\right|}_{\l_j=\l_i}
+ \left(\tilde
W(\l_i)\right)^2+
\frac{1}{n}\sum_{j\neq i}\frac{\tilde W(\l_j)}{\l_j-\l_i}=
\frac{\l_i^2}{4}-\baA +\half
\label{SD}
\eeq
where
\be
\tilde
W(\l_i)=W(\l_i)+\frac{\l_i}{2}+\frac{1}{2n}\sum_{j\neq i}\frac{1}{\l_i-\l_j}
\label{tildeW}
\ee
\eq{SD} can be obtained by
inserting \eq{rho} into the loop equation~\rf{loop}.

Substituting Eqs.~\rf{eps},~\rf{5.20} and~\rf{2loop} and assuming
multiplicative renormalizability of the higher genera contributions to
the correlators (as in the double scaling limit) one gets from
equation~\rf{SD} the following equation
\be
{\left.\frac{1}{n^2}
\frac{\d \tilde W^{kont}(m_i^2)}{\d m_j^2}\right|}_{m_j^2=m_i^2}+
\left(\tilde W^{kont}(m_i^2)\right)^2
+\frac{1}{n}\sum_{j\neq i}
\frac{\tilde{W}^{kont}(m^2_j)}{m^2_j-m^2_i}=
\frac{m_i^2}{4}
\label{SDkont}
\ee
where
\be
\tilde W^{kont}(m^2_i)= W^{kont}(m^2_i)+\frac{m_i}{2}
+\frac{1}{2n}\sum_{j\neq i}
\frac{m_j}{m_i}\left(\frac{1}{m_i^2-m_j^2}\right)
\ee
This is exactly the Schwinger-Dyson equation of the Kontsevich model.

\subsection{The equivalence of the d.s.l.\ with
Kontsevich model \label{lkontsevich}}

The equivalence of the Kontsevich model and \dsl\ of \hm\ in genus zero
was demonstrated in Ref.~\cite{MS} by comparing explicit solutions.
In order to compare next orders of the genus expansions, let us consider
the following
expression for $F_g^{Kont}$
 conjectured in Ref.~\cite{IZ} for the Kontsevich model:
\beq
F_g^{Kont} = \sum_{\a_j >1} \la \a_1 \ldots {\a_s}|\a \RA_g^{kont}
\frac{I_{\a_1}\cdots I_{\a_s}}{(I_1-1)^\a}  \hspace{1.7cm}g\geq1
\label{add3}
\eeq
where  the moments $I_k$'s  are defined by%
\footnote{This definition differs by a factor $-(2k-1)!!$
from the $I_k$'s used in Ref.~\cite{IZ}.}:
\beq I_k(M) = \frac{1}{n} \sum_{j=1}^n \frac{1}{(m_j^2-2u_0)^{k+1/2}}
\hspace{1.7cm} k\geq 0\label{add4}
\eeq
and $u_0(M)$ is determined by the equation $u_0 = I_0 (u_0,M)$, \ie
by \eq{add5}.
When we compare the coefficients $\la \cdot \RA^{kont}_g$
calculated in Ref.~\cite{IZ} with the $\la \cdot \RA^{herm}_g$
calculated in \dsl\ in Section~\ref{F_g}, we see that they are
{\it identical\/}. In this section we show that this is true to
all orders in the genus.

For any value of $\eps$ in \rf{add7} we have the expansion \rf{fgen} of
$F_g$ of the hermitian matrix model in terms of $M_k$'s and
$J_k$'s. We are now going to show that the terms which survive
in the limit $\eps \to 0$ are the same as those that
survive in the double
scaling limit \rf{fgendsl}, and that the $M_k$'s are related directly
to the $I_k$'s of the Kontsevich model defined by~\rf{add4} and~\rf{add5}.

Let us now discuss the scaling behaviour of
the  moments $M_k$ and $J_k$.
By substituting (\ref{rho}) into Eqs.~(\ref{moment1}) and
 (\ref{moment2}), we obtain
for the moments $M_k$, $J_k$:
\bea
M_k&=&\frac 1n \sum_j  \frac{1}{(\l_j-x)^{k+1/2}(\l_j-y)^{1/2}}
-\delta_{k1} \hspace{1.5cm}{k\geq1}\;,\label{newmoment1}\\
J_k&=&\frac 1n \sum_j  \frac{1}{(\l_j-x)^{1/2}(\l_j-y)^{k+1/2}}
-\delta_{k1} \hspace{1.5cm}{k\geq1}\;.
\label{newmoment2}
\eea
 From this representation and Eqs.~\rf{eps}, \rf{epsxy} it is easy to
infer for the moments $M_k$ and $J_k$  and for
the parameter $d$ the following scaling rules:
\bea
J_k&{}&\to -2^{-(3k/2+1)}\eps^{(3k+1)/2}I_0+\delta_{k1},\label{s1}\\
M_k&{}&\to -2^{(k-1)/2}\eps^{-(k-1)/2}(I_k-\delta_{k1}),\label{s2}\\
d&{}&\to 2^{3/2}\eps^{-3/2}.\label{s3}
\eea
where $I_k$'s  are the the Kontsevich moments defined by \rf{add4}, \rf{add5}.

Let us turn to the general expression (\ref{fgen}) for $F_g$. Then the
power in $\eps$ for a specific term
\beq
\frac{M_{\a_1} \cdots M_{\a_s} J_{\b_i} \cdots J_{\b_l}}
{d^{\g} M_1^\a J_1^\beta}
\label{add8}
\eeq
is
\beq
[\eps]=\sum_{i=1}^l{3\b_i+1\over 2}-\sum_{j=1}^s{\a_j-1\over 2}+\frac 32\g.
\eeq
Clearly, due to the formulas (\ref{inv2F}) and (\ref{boundg}), we
find that $[\eps]\geq 0$. $[\eps]=0$ is
 possible if and only if $l=0$, $\g=g-1$ and
\beq
\sum_{j=1}^s(\a_j-1)=3g-3.\label{s4}
\eeq
This means that we have precisely the same terms surviving as in the
generic  double scaling limit \rf{fgendsl}.
By taking a closer look at the loop insertion operator~\rf{loopins} one
finds that such terms always have $\beta=0$ and hence $N_M=2-2g$. We
therefore conclude that starting from the generic hermitian matrix model
and letting $\eps\ra 0$ we will reproduce
\rf{add3}, \rf{add4} and \rf{add5} the
coefficients in front of the $I_k$ terms being those inherited from
the hermitian matrix model. Hence we have
$\la \cdot \RA_g^{herm}=\la \cdot \RA_g^{Kont}$
 since we saw explicitly that the partition function
$Z[g,N(\baA )]$ is identical to the partition function of the
Kontsevich model in the  limit $\eps \to 0$, $\baA =1/{2}\eps^3$.

Thus the coefficients $\la \cdot \RA_g^{herm}$ which survive in the
double scaling limit have  an interpretation as intersection indices on
moduli space, since this is the interpretation of the coefficients of
the Kontsevich model. In Section~\ref{disk} we are going to show that
the additional coefficients which appear away from the double scaling limit
can also be given a geometric interpretation on moduli space.

\subsection{Intersection indices on  moduli space \label{i.i.}}

It has been known for some time that matrix models might be very useful
in providing explicit realizations of the moduli spaces $\cM_{g,s}$ of
Riemann surfaces of genus $g$ and $s$ marked points $x_1,\ldots,x_s$.
The first example
is the Penner model \cite{Pen}, which allowed a relatively simple
calculation of the virtual Euler characteristics of these spaces.
The next example is the Kontsevich model, which allows a
calculation of intersection indices on $\cM_{g,s}$.

It is not our intention here to define or discuss these concepts in any
detail. Let us only note, to fix the notation, that there exist $s$ natural
line bundles $\cL_i$, $i=1,...,s$ on $\cM_{g,s}$. The fiber of $\cL_i$
at a point
$\Sigma \in \cM_{g,s}$ is the cotangent space to the point $x_i$, viewed
as a point on the surface $\Sigma$. The line bundles have  first Chern
classes $c_1(\cL_i)$, which can be represented by the curvature 2-form
of an arbitrary $U(1)$ connection on $\cL_i$. If we have
non-negative integers $\a_i$ such that
\beq
\sum_{i=1}^s \a_i = \oh {\rm real~dim}\; \cM_{g,s} = 3g-3+s \label{add10}
\eeq
it is possible to form the integral~\cite{W1}
\beq
\LA \tau_{\a_1} \cdots \tau_{\a_s} \RA =
\int_{\bar{\cM}_{g,s}} c_1(\cL_1)^{\a_1}\cdots c_1(\cL_s)^{\a_s}
\label{add11}
\eeq
where products and powers are exterior products and powers and
$\bar{\cM}_{g,s}$ refers to a suitable compactification (the so-called
Deligne--Mumford compactification) of $\cM_{g,s}$.

The numbers $\LA \tau_{\a_1} \cdots \tau_{\a_s} \RA$ are topological
invariants and one of the achievements of Kontsevich was that he
found~\cite{Kon91} a more manageable representation by using a combinatorial
decomposition of $\cM_{g,s}$ inherited from the physicists ``fat-graph''
expansion of the hermitian matrix integrals. If we have a double-line
graph with $s$ faces (dual to the $s$ punctures), we can assign a  perimeter
$p_f=\sum_e l_e$ to each face, $l_e$ being the length of the edge $e$
and the sum being over edges
which constitute the face $f$. The set of such double-line graphs
with assigned $l$'s is a combinatorial model $\cM^{comb}_{g,s}$ of
moduli space. Cells $G$ have dimensions over the real numbers obtained
by counting the number
of independent $l_e$'s for fixed $p_f$'s. The result is that
\beq
{\rm real~ dim} \;G \leq 2(3g-3+s) \label{add12}
\eeq
where the equality sign is valid if and only if all vertices of the graph
$G$ are trivalent.

For each face $f_i$ one can introduce the 2-form $\om_i$, which can be
considered as the combinatorial version of $c_1 (\cL_i)$ related to the
puncture $x_i$:
\beq
\om_i = \sum_{a < b} d\left( \frac{l_a}{p_{f_i}}\right) \wedge
d\left(\frac{l_b}{p_{f_i}}\right) \label{add13}
\eeq
where $l_a$ are the lengths of the edges of $f_i$, which are assumed to
be cyclic ordered. For  fixed $p_{f}$'s
we can now write the formula for intersection indices as
\beq
\LA \tau_{\a_1} \cdots \tau_{\a_s} \RA = \int \om_1^{\a_1}\cdots \om^{\a_s}_s
\label{add14}
\eeq
provided the integral is over cells $G$ of maximal dimension,
i.e. {\it provided all vertices are trivalent}. Let us define denote
the complex dimension of $\cM_{g,s}$ by $d_{g,s}$, \ie
\be
d_{g,s}=3g-3+s\,,
\ee
 and introduce
the 2-form
\beq
\Omega = \sum_{i=1}^{s} p_{f_i}^2 \om_i.
 \label{add15}
\eeq
It acts as a generating function for intersection indices:
\beq
\int \Omega^{d_{g,s}}= \sum_{\a_1+ \ldots + \a_s = d_{g,s}}
\LA \tau_{\a_1} \cdots \tau_{\a_s} \RA_g \;
 d_{g,s}! \prod_{i=1}^s \frac{p_{f_i}^{2\a_i}}{\a_i !}
\label{add16}
\eeq
By a Laplace transformation we can trade the $p_i$'s for $m_i$'s by
\beq
\int^\infty_0 dp_i \;e^{-p_i m_i} \; p^{2\a_i} = (2\a_i)! \; m_i^{-2\a_i-1}
\label{add17}
\eeq
and one of the main results of Kontsevich is that
\beq
\sum_{\a_1+ \ldots +\a_s = d_{g,s}}
\LA \tau_{\a_1} \cdots \tau_{\a_s} \RA_g
\prod_{i=1}^s \frac{(2\a_i-1)!!}{m_i^{2\a_i+1}}       =
\sum_{G} \frac{2^{-V}}{Aut(G)} \prod_{e \in G} \frac{2}{m_f+m_{f'}}
\label{add18}
\eeq
where the summation on the right hand side is over all double-line,
trivalent, connected graphs $G$ of genus $g$ and with $s$ faces.
The quantity denoted by
$V$ is the number of vertices in $G$, $e$ denote an edge in $G$
and $f$ and $f'$ the faces sharing a given double-line $e$.
Finally $Aut (G)$ is the order of symmetry group of $G$.
For future use we note that that the last factor can be written as
\beq
\prod_{e \in G} \frac{1}{m_f+m_{f'}} =
\int_0^\infty  \prod_{e \in G}dl_e \; e^{-l_e(m_f+m_{f'})}
\label{add19}
\eeq

When we compare the right hand side of \eq{add18} with the definition of
the Kontsevich integral \rf{Kont} we see immediately that
\rf{add18} represents the sum of the connected diagrams of
genus $g$ and $s$ faces which would be generated by a perturbative
expansion with respect to the gaussian part of the action.
In the next section we shall extend this result to a more general
interaction.

\subsection{The discretized moduli space \label{disk}}

As we shall see
the results obtained in this article {\it away} from the critical regime
is related to a representation of moduli space which incorporates
explicitly the boundary components of the Deligne--Mumford
compactification (or reduction) procedure.
The basic ingredient in the new construction
is a {\it discretization} of $\cM^{comb}_{g,s}$. The length
$l_e$ of an edge is assumed the belong to $ \eps Z_+$, where $ \eps$
is an expansion parameter, which eventually is going to be associated
with the one in the Section~\ref{lkontsevich}.
This means that the Laplace transform in $p_i$ performed in \rf{add17}
is going to be replaced by a sum. The effect of this summation is
heuristically described by referring to the representation \rf{add19}:
\beq
\int_0^\infty  \prod_{e \in G}dl_e \; e^{-l_e(m_f+m_{f'})} \to
\eps^{3d_{g,s}} \prod_{e \in G} \sum_{n_e=1}^\infty e^{-\eps n_{e} (m_f+m_f')}
= \eps^{3d_{g,s}} \prod_{(ij)} \frac{1}{e^{\eps (m_i+m_j)} -1}
\label{add20}
\eeq
where the power of $\eps$ is valid for connected trivalent graphs.
In the last product we recognize the propagator for the Kontsevich--Penner
model \rf{kp1} which involves the factor
\beq
\frac{1}{\h_i \h_j -1}.   \label{add21}
\eeq
provided we make the assignment $\h_i = \exp(\eps m_i)$. This assignment
is nothing but the one already made in \rf{add7}, and the $\eps$ assignment
in \rf{add7} is also consistent with the $\eps$ factor  in \rf{add20}.
In order to fit our notations to the ones by Kontsevich it will be sometimes
conveninent to use $\tilde\baA = 2\baA =\eps^{-3}$.
In fact the perturbative expansion of the Kontsevich--Penner model
\rf{kp1} can be written as
\beq
\ln Z_{KP}[\h,n] =
\sum_{g=0}^{\infty}\sum_{s=0}^{\infty} n^{2-2g}\left( \tilde\baA
\right)^{2-2g-s}
\sum_{G} \frac{2^{-V}}{Aut (G)} \prod_{(ij)}\frac{2}{\h_i\h_j -1}
\label{add22}
\eeq
where the notation is as in \rf{add18}, {\it except} that  $\sum_G$ denotes
the summation over {\it all} connected double-line graphs of genus $g$
and $s$ faces, not only the trivalent graphs as in \rf{add18}.
In this way we see that the Kontsevich--Penner model for finite $\eps$
provides a generalization of the Kontsevich model in the sense that
it in the limit $\eps \to 0$ just reduces to the Kontsevich model,
while it for $\eps > 0$ describes a discretized version of moduli
space, which however allows us to access the boundary in more detail.

By use of this formalism it can be shown that the Kontsevich--Penner
model allows an expansion, related to the Deligne--Mumford
reduction in moduli space, which can be viewed as a generalization
of the corresponding expansion of the Kontsevich model~\cite{Ch}.
For the coefficient of $n^{2-2g}\baA^{2-2g-s}$
in the expansion~\rf{add22} of $\ln Z_{KP}$ one has

\bea
F_{g,s}&=&\eps^{6g-6+3s}\sum_{reductions} (-1)^{r_p}
\frac{(d_{g,s}-r_p)!}{d_{g,s}!}
\nonumber\\
&{}&\times \prod_{j=1}^p
\left[ \sum_{\sum \a_a=d_j}\LA \tau_{\a_1}\dots\tau_{\a_{f_j}}
\tau_0^{(1)}\dots
\tau_0^{(k_j)}\RA_{g_j}\right.\frac{(f_j)!}{(f_j+k_j)!}\nonumber\\
&{}&\times \left.\str\prod_{k=1}^{f_j}
\left(
\frac{1}{\eps}\frac{\partial}{\partial m_k}\right)^{2\a_k}
\,\left.\frac{(-1)^{f_j}}{\prod_{k=1}^{f_j}(\e ^{\eps m_k}-1)}
\right|_{{symmetrized\atop
\e^{\eps m_k}\to -\e^{\eps m_k}}} \right]. \label{red1}
\eea
As opposed to what was the case for the Kontsevich model one should now
take into account all reductions of the original Riemann surface.
The notation is as follows. For a given reduction, $p$ denotes the number
of connected components of the reduced Riemann surface. These components
are labeled by the index $j$. The $j$'th component has genus $g_j$,
$f_j$ original punctures and $k_j$ punctures resulting from the
reduction procedure. Note that there is no $m_k$ variables corresponding to
these
new insertions, that leads to appearing of additional symmetrical factors
$(f_j)!/(f_j+k_j)!$ in \rf{red1}.
The corresponding moduli space has complex
dimension $d_j=3g_j-3+f_j+k_j$. The quantity $r_p$ is the power of
reduction, defined by $\sum_{j=1}^p d_j=d-r_p$.
The symmetrization ensures that $F_g$ is invariant under the replacements
$M_k\to (-1)^{k+1}J_k$ as it must be.

The expression \rf{red1}
resembles in many details the original answer by Kontsevich, the
only two differences being that firstly the variables of expansion, which in
the
Kontsevich case appear in the combination
$-(2\a_i-1)!!\str m^{-2\a_i-1}$, in \rf{red1}  appear in
a more involved expression. Secondly the complete answer
(\ref{red1}) contains information about the
reductions of the Riemann surface.

Since, as shown explicitly in Section~\ref{l.p.}, the
Kontsevich--Penner model can be mapped onto the general
hermitian matrix model the information about this reduction is encoded in
our expansion of the hermitian matrix model in terms of the moments $M_k,J_k$,
 and by exploring it we may hope to
gain some information
about the boundary of the moduli space.
In order to do this, we need to expand moments
$M_k$, $J_k$ and $d$ in terms of the somewhat unusual expansion parameters
standing on
the right hand side of Eq.~(\ref{red1}).
As the first step we need to
resolve  the constraint equations
(\ref{xandy}) in terms of these parameters. As the first
approximation we have (in terms of eigenvalues):
\bea
M_k&{}&\sim -\frac{1}{\sqrt{\baA}^{k+1}}\, \frac 1n \sum _{i=1}^n
\frac{\e ^{\eps m_i (k+1)}}{(1-\e^{\eps m_i})^{2k+1}(1+\e^{\eps m_i})}
\,+\delta_{k1},\\
J_k&{}&\sim -\frac {1}{\sqrt{\baA}^{k+1}}\,\frac 1n \sum_{i=1}^n \frac
{\e ^{\eps m_i(k+1)}}{(1-\e^{\eps m_i})(1+\e^{\eps m_i})^{2k+1}}\,+
\delta_{k1},\\
d &{}&\sim \sqrt{\baA}\left\{
4-\frac{1}{\baA n}
\sum _{i=1}^n \frac{2}
{(\e^{\eps m_i}-1)(1+\e^{\eps m_i})}\right\}.
\eea

Let us now check the validity of the expansion (\ref{red1}), for the simplest
non-trivial example, the moduli space $F_{1,1}$
(a torus with one marked point). Since
$s=1$, we should extract only single trace terms
from the formal expansion of the answer in product of traces.
 Using the explicit answer for $F_1$, we find that
\bea
\left.-\frac {1}{24}\ln d^4J_1M_1\right|_{F_{1,1}} &=&
 -\frac {1}{24\xi}\str\left\{
-\frac{2}{(\e^{\eps m}+1)(\e^{\eps m}-1)}\right.\nonumber\\
&{}&\left.-\frac{\e^{2\eps m}}{(\e^{\eps m}+1)^3(1-\e^{\eps m})}
-\frac{\e^{2\eps m}}{(\e^{\eps m}+1)(1-\e^{\eps m})^3}\right\}.
\label{f11}
\eea
Doing this sum we reconstruct the answer (which can be also obtained
independently
using diagram technique):
\bea
F_{1,1}&=&-\frac{1}{12\xi}\str\,\frac{3\e^{2\eps m}-1}{(\e^{2\eps m}-1)^3}
\nonumber \\
&= &
\left.\str\left\{-\frac{1}{24\tilde\xi}
 \,\frac{\partial ^2}{\partial m^2}\,\frac{1}{\eps ^2}\,
\frac {1}{\e^{\eps m}-1}\,+\frac {1}{6\tilde\xi}\,\frac{1}{\e^{\eps
m}-1}\right\}
\right|_{{symmetrized\atop\e^{\eps m_k}\to -\e^{\eps m_k}}},
\label{f11expan}
\eea
where we again substitute $\tilde\xi =2\xi =\eps^{-3}$ in order to compare with
the
Kontsevich model answers.
That is, the form  (\ref{red1}) is explicitly reproduced. The first
coefficient,
$\frac {1}{24}$, is just the Kontsevich index, and the second term corresponds
to the
unique reduction of the torus, \ie the sphere with an additional factor $1/3!$
due to
reduction.

This rather simple, but instructive example shows that \hm\ might be
able to offer
some new insight in the fine structure of moduli space.

\newsection{Discussion}

We would like to mention at the end one more approach to \hm\
which is based
on its interpretation as an integrable system.
The appropriate hierarchy for \hm\ is the Toda-chain
one~\cite{Toda,KMMM} and its
partition function can be represented as the corresponding
$\tau$-function. This $\tau$-functin
obeys~\cite{dvir} a set of discrete
Virasoro constraints quite similarly to the
continuum  partition function~\cite{DVV}.
The relation between these two sets of the Virasoro constraints was studied
in Ref.~\cite{MMMM}.

As is discussed in Ref.~\cite{MMMM}, these Virasoro constraints for \hm\ are
nothing but the loop equation~\rf{loop}. Therefore, the explicit solutions
found in the present paper are simultaneously solutions of the Virasoro
 constraints. We do not refer, however, to the fact that the partition
function~\rf{partition} is the $\tau$-function.
In the double scaling limit this relation
gives a powerful computational method to obtain higher orders of
the genus expansion based on the
Gelfand--Dikii technique.
Its analogue for \hm\ away from the double scaling limit
is yet missing. We think that our explicit calculations prompt that such
a technique should exist.

\vspace{12pt}

\noindent
{\bf Acknowledgements} \hspace{0.5cm} We thank A.~Mironov for
interesting  and  valuable discussions.
L. Ch.\
would like to thank for hospitality Professor
P.K.Mitter and Laboratoire de
Physique Th\'eorique et des Hautes \'Energies.
Yu.\ M. thanks the theoretical physics department of UAM for the
hospitality at Madrid.
\vspace{12pt}

\setcounter{section}{0}

\appendix{The d.s.l.\ of complex matrix model}

The partition function of the complex matrix model and the definition of
its correlators were given in Section~\ref{complex}. In the same way as
 for the hermitian one matrix model, one can obtain the correlators
from the free energy by application of the loop insertion operator,
$\frac{d}{dV_C(p)}$:
\beq
W^C(p_1,\ldots, p_s)=\frac{d}{dV_C(p_s)}\;\frac{d}{dV_C(p_{s-1})}\ldots
\frac{dF^C}{dV_{C}(p_1)}
\eeq
where
\beq
\frac{d}{dV_C(p)}\equiv
-\sum_{j=1}^{\infty}\frac{j}{p^{2j+1}}\frac{d}{dg_{j}}\,.
\eeq

In Ref.~\cite{complex} an iterative scheme for calculating
$W_g^C(p)$ for any genus $g$ starting from $W_0^C(p)$ was developed. The
idea was very much the same as described in Section~\ref{Wg}. The basic
ingredient was the loop equation for the complex matrix model:
\be
\cic W^C(\om)=[W^C(p)]^2+\frac{1}{N^2}W^C(p,p)\,,
\label{cloop}
\ee
which after insertion of the genus expansion of the correlators reads
\beq
\left\{\hat{K}_C-2W^C_{0}(p)\right\}W^C_g(p)=\sum_{g'=1}^{g-1}
W^C_{g'}(p)\;W^C_{g-g'}(p)
+\frac{d}{d V_C(p)}W^C_{g-1}(p)\,. \label{loopc}
\eeq
The operator $\hat{K}_C$ is given by
\beq
\hat{K}_Cf(p)\equiv \cic f(\om)
\eeq
where $V_C(\om)=\sum_j g_j\om^{2j}/j$ and $P$ is a path which encloses the
singularities of $W_C(p)$. It was assumed that these consisted of only
one cut $[-\sqrt{z},\sqrt{z}]$ on the real axis which in the eigenvalue
picture corresponds to assuming that the eigenvalue density for
$\phi^{\dg}\phi$ has support only on the interval $[0,z]$.

To characterize the matrix model potential, instead of the
couplings $g_j$, the following moments were introduced
\bea
N_{k}&=&\cimc \frac{V_C'(\om)}{w^{2k+1}(w^2+c)^{1/2}} \hspace{1.0cm} k\geq
0\;,\\
I_{k}&=&\cimc \frac{\om V_C'(\om)}{(w^2+c)^{k+1/2}}  \hspace{1.6cm} k\geq 0\;.
\eea
For the complex matrix model the $m^{th}$ multi-critical point is
characterized by the eigenvalue density for $\phi^{\dg}\phi$ having a
zero of order $(m-1)$ at the endpoint $z$ of its support. It is easily
shown that the condition for being at this point is
$I_1 = I_2 =\ldots =I_{m-1}=0$ and  $I_m \neq 0$.
In Ref.~\cite{complex} it was proven by induction that $W_g^C(p)$
can be written as
\beq
W^C_g(p)=\sum_{n=1}^{3g-1}{\cal A}_g^{(n)}X^{(n)}(p)
       +\sum_{m=1}^g {\cal D}_g^{(m)} Y^{(m)}(p)\,.
\label{conjectureC}
\eeq
The basis vectors $X^{(n)}(p)$ and
$Y^{(n)}(p)$ fulfill
\bea
\left\{\hat{K}_C-2W^C_0(p)\right\}X^{(n)}(p) &= & \frac{1}{(p^2-z)^n}
\,,\\
\left\{\hat{K}_C-2W^C_0(p)\right\}Y^{(n)}(p) &=&\frac{1}{p^{2n}}
\eea
and are given by
\bea
X^{(n)}(p)&=&\frac{1}{I_1}\left\{\Phi^{(n)}(p)
         -\sum_{k=1}^{n-1}X^{(k)}(p)I_{n-k+1}
          \right\}\,,
\label{X}\\
Y^{(n)}(p)&=&\frac{1}{N_0}\left\{\Omega^{(n)}(p)
         -\sum_{k=1}^{n-1}Y^{(k)}(p)N_{n-k}
          \right\}
\label{Y}
\eea
where
\bea
\Phi^{(n)}(p)&=&\frac{1}{(p^2-z)^{n+1/2}}\,,\label{Phi}\\
\Omega^{(n)}(p)&=&\frac{1}{p^{2n}(p^2-z)^{1/2}}\,.\label{Omega}
\eea

The iteration process for determining $W^C_g(p)$ is quite similar to
the hermitian matrix model. To start one needs an expression for
$W^C_0(p,p)$. Using a trick similar to~\rf{trick}, one finds
\beq
W^C_0(p,p)=\frac{z^2}{16p^2(p^2-z)^2}\,.
\eeq
In the general step one makes use of the following rewriting of the loop
insertion operator
\beq
\frac{d}{dV_C(p)} = \sum_{i}\frac{dI_i}{dV_C(p)}\frac{\partial}{\partial I_i}
+\sum_{j}\frac{dN_j}{dV_C(p)}\frac{\partial}{\partial N_j}
+\frac{dz}{dV_C(p)} \frac{\partial}{\partial z}
\label{loopio}
\eeq
where
\bea
\frac{dI_i}{dV_C(p)}&=&-i\Phi^{(i)}(p)
-(i+1/2)\left\{\Phi^{i+1}(p)-\frac{I_{i+1}}{I_1}\Phi^{(1)}(p)\right\}z \,,
\label{dIdV}
\\
\frac{dN_j}{dV_C(p)}&=&\frac{1}{2}\left\{
-(2j+1)\Omega^{(j+1)}(p)-\frac{1}{p^{2j}(p^2+c)^{3/2}}\right\}\nonumber
\\
&&-\frac{1}{2}\left\{\frac{1}{c}\sum_{l=0}^{j}N_{j-l}\frac{(-1)^l}{c^l}
   +\frac{(-1)^{j+1}}{c^{j+1}}I_1\right\}
   \frac{dz}{dV_C(p)} \,,
\label{dNdV}
\\
\frac{dz}{dV_C(p)}&=&\frac{z}{I_1}\phi^{(1)}(p) \,.\label{dzdV}
\eea
Calculating the right hand side of the loop equation one gets an
expression involving fractions of the type $p^{-2n}(p^2-z)^{-m}$.
Decomposing these fractions into fractions of the type $p^{-2i}$,
$(p^2-z)^{-j}$ allows one to identify immediately the coefficients
${\cal A}_g^{(n)}$ and ${\cal D}_g^{(n)}$.

Let us now modify the iterative procedure so that it gives us directly
the result in the double scaling limit. The double scaling limit of the
correlators for the $m^{th}$ multi-critical model is obtained by fixing
the ratio of any given coupling and the first one to its critical value
and scaling $p^2$ and $z$ as in~\rf{scalps} and~\rf{scalzs}. It is easy
to convince oneself that, just as it was the case for the hermitian one
matrix model, adjusting the iterative procedure to give only the double
scaling relevant results amounts to starting from the double scaled
version of $W^C_0(p,p)$ and to discarding all operators in $d/dV_C(p)$ which
do not lower the power of $a$ by $m+3/2$. One finds
\beq
W^C_0(p,p) =\frac{z_c}{16(p^2-z)^2} \hspace{1.0cm} (d.s.l.)
\label{dsl1}
\eeq
and the loop insertion operator reduces to
\beq
\frac{d}{dV_C(p)} = \sum_{i}\frac{dI_i}{dV_C(p)}\frac{\partial}{\partial I_i}
+\frac{dz}{dV_C(p)} \frac{\partial}{\partial z}
\hspace{1.0cm}(d.s.l.)
\label{dsl2}
\eeq
where
\bea
\frac{dI_i}{dV_C(p)}&=&
-(i+1/2)\left\{\Phi^{i+1}(p)
-\frac{I_{i+1}}{I_1}\Phi^{(1)}(p)\right\}z_c
\hspace{1.0cm} (d.s.l) \,,\label{dsl3}
\\
\frac{dz}{dV_C(p)}&=&\frac{z_c}{I_1}\Phi^{(1)}(p)\hspace{1.0cm} (d.s.l.) \,.
\label{dsl4}
\eea

Let us compare the formulas~\rf{dsl1} -- \rf{dsl4} with the corresponding
formulas for the double scaling limit of the symmetric hermitian one
matrix model~\rf{Hdsl1} -- \rf{Hdsl4}. Noticing that $d_c^2=4z_c$
and that in the double scaling limit
\beq
M_k=d_c^{k-1}I_k\,,
\eeq
we see that in this limit $W_0^{C}(p,p)=1/4\cdot W_0^{(S)}(p,p)$,
$d/dV_C(p)=1/4\cdot d/dV_S(p)$ and $X^{(n)}(p)=\tilde{\chi}^{(n)}(p)$.
Since the loop equation looks the same for the complex and hermitian
matrix models (cf.\ Eqs.~\rf{loopg} and~\rf{loopc}), we thus have
\beq
W_g^{C}(p_1,\ldots,p_s) = \frac{1}{4^{g+s-1}}W_g^{(S)}(p_1,\ldots,p_s)
\hspace{0.7cm}(d.s.l.)\,.
\eeq
Furthermore, since the generating functional is related to the free
energy in the same way for the two models, we also have
\beq
F_g^{C}= \frac{1}{4^{g-1}}F_g^{(S)}
\hspace{0.7cm}(d.s.l.)\,.
\eeq
Hence, the hermitian and  complex one matrix models are equivalent in the
double scaling limit. However, away from the double scaling, we clearly
have to distinct models. We draw the attention of the reader to the fact
that no assumptions about
the type of critical behaviour entered the arguments of this
appendix.

\end{document}